\documentclass[american,sigconf,nonacm]{acmart}

\newcommand\vldbdoi{XX.XX/XXX.XX}
\newcommand\vldbpages{XXX-XXX}
\newcommand\vldbvolume{14}
\newcommand\vldbissue{1}
\newcommand\vldbyear{2020}
\newcommand\vldbauthors{\authors}
\newcommand\vldbtitle{\shorttitle}

\newcommand\vldbpagestyle{plain}

\usepackage[hide]{todo}
\usepackage{dirtytalk}
\usepackage{appendix}
\usepackage{algorithmicx}
\usepackage{algpseudocode}
\usepackage{siunitx}
\usepackage[normalem]{ulem}

\usepackage[utf8]{inputenc}         
\usepackage[babel]{csquotes}        
\usepackage[plain]{fancyref}        
\usepackage[T1]{fontenc}
\usepackage{algorithm}
\usepackage{algpseudocode}          
\usepackage{amsmath}
\usepackage{babel}                  
\usepackage{float}                  
\usepackage{hyperref}               
\usepackage{isodate}                
\usepackage{listings}
\usepackage{lstautogobble}          
\usepackage{multicol}               
\usepackage{subcaption}             
\usepackage{subscript}              
\usepackage{tikz}
\usepackage{wasysym}                
\usepackage{xspace}                 

\usetikzlibrary{
    arrows,
    arrows.meta,
    backgrounds,
    calc,
    chains,
    decorations,
    decorations.markings,
    decorations.pathreplacing,
    fit,
    matrix,
    patterns,
    positioning,
    scopes,
    shadows,
    shapes,
}

\theoremstyle{definition}

\AtBeginDocument{

}

\newfloat{lstfloat}{htbp}{lop}
\floatname{lstfloat}{Listing}

\definecolor{c_mutable}{RGB}{32, 167, 178}
\definecolor{c_bg}{RGB}{255, 243, 230}
\definecolor{OliveGreen}{HTML}{3b7f31}
\definecolor{LimeGreen}{HTML}{8cc63e}
\definecolor{Turquoise}{HTML}{00b4cd}
\definecolor{ProcessBlue}{HTML}{00afef}
\definecolor{PineGreen}{HTML}{008a72}
\definecolor{Dandelion}{HTML}{fdbb42}

\newcommand{\noun}[1]{\textsc{#1}\xspace}

\newcommand{\ca}[1]{\citeauthor{#1}~\cite{#1}}

\newenvironment{abstract*} {\noindent{\bfseries Abstract ---}} {}

\newcommand{\Wasm}{\noun{WebAssembly}}
\newcommand{\llvm}{\noun{LLVM}}

\newcommand{\mutable}{\mbox{mu\hspace{-.5pt}\textit{\color{c_mutable!90!black}t}\hspace{.5pt}able}\xspace}
\newcommand{\hyper}{\mbox{\noun{HyPer}}\xspace}
\newcommand{\duckdb}{\mbox{\noun{DuckDB}}\xspace}
\newcommand{\postgres}{\mbox{\noun{PostgreSQL}}\xspace}
\newcommand{\Cpp}{C\texttt{++}\xspace}

\usetikzlibrary{fit}
\pgfdeclarelayer{bg}
\pgfsetlayers{bg,main}

\tikzstyle{block}=[rectangle,thick,draw]
\tikzstyle{rel}=[circle,thick,draw,inner sep=1mm]
\tikzstyle{ann}=[font=\fontsize{6}{7.2}\selectfont]

\tikzstyle{next}=[->,>=stealth',thick,shorten <=2pt,shorten >=2pt]
\tikzstyle{knows}=[->,>=stealth,dashed,shorten <=1pt,shorten >=1pt]
\tikzstyle{is}=[->,>=triangle 60,thick,shorten <=2pt,shorten >=2pt]
\tikzstyle{owns}=[{diamond}-,thick]
\tikzstyle{has}=[{open diamond}-,thick,fill=white]

\tikzset{
    font=\fontsize{8}{9.6}\selectfont,
    invisible/.style={opacity=0},
    visible on/.style={alt={#1{}{invisible}}},
    alt/.code args={<#1>#2#3}{%
        \alt<#1>{\pgfkeysalso{#2}}{\pgfkeysalso{#3}} 
    },
}

\lstdefinelanguage{WebAssembly}{
  sensitive=true,
  otherkeywords={},
  morekeywords=[1]{i32, f32, i64, f64},
  keywordstyle={[1]\color{RedViolet}},
  morekeywords=[2]{0},
  keywordstyle={[2]\color{RedViolet}},
  morekeywords=[3]{add, const}
  keywordstyle={[3]\color{RoyalBlue}},
  morekeywords=[4]{},
  keywordstyle={[4]\color{Rhodamine}},
  morekeywords=[5]{module, func, param, result, global, get_global, mut, set_global, export, import, memory, data, get_local, set_local, elem, table, call, call_indirect, type, select},
  keywordstyle={[5]\color{ProcessBlue}},
  morekeywords=[6]{=,;},
  keywordstyle={[6]\color{SeaGreen}},
  morekeywords=[7]{(,),[,],.},
  keywordstyle={[7]\color{black}},
  numberstyle=\tiny\color{black},
  rulecolor=\color{black},
  morecomment=**[l][\itshape\color{OliveGreen}]{;;},
  morecomment=[s]{(;}{;)},
  commentstyle=\color{OliveGreen}\itshape
}

\algnewcommand\algorithmicswitch{\textbf{switch}}
\algnewcommand\algorithmiccase{\textbf{case}}
\algdef{SE}[SWITCH]{Switch}{EndSwitch}[1]{\algorithmicswitch\ #1\ \algorithmicdo}{\algorithmicend\ \algorithmicswitch}%
\algdef{SE}[CASE]{Case}{EndCase}[1]{\algorithmiccase\ #1}{\algorithmicend\ \algorithmiccase}%
\algtext*{EndCase}%

\algrenewcommand\algorithmicrequire{\textbf{Input:}}
\algrenewcommand\algorithmicensure{\textbf{Output:}}

\colorlet{AlgoComment}{LimeGreen!80!black}

\makeatletter
\algrenewcommand\ALG@beginalgorithmic{\fontsize{7}{7}\selectfont}
\makeatother

\begin{document}

\title{Fast Compilation and Execution of SQL Queries with WebAssembly}

\author{Immanuel~Haffner}
\affiliation{%
    \institution{Saarland Informatics Campus}
}
\email{immanuel.haffner@bigdata.uni-saarland.de}

\author{Jens~Dittrich}
\affiliation{%
    \institution{Saarland Informatics Campus}
}
\email{jens.dittrich@bigdata.uni-saarland.de}


\begin{abstract}

    Interpreted execution of queries, as in the vectorized model, suffers from interpretation overheads.  By compiling
    queries this interpretation overhead is eliminated at the cost of a compilation phase that delays execution,
    sacrificing latency for throughput.  For short-lived queries, minimizing latency is important, while for
    long-running queries throughput outweighs latency.  Because neither a purely interpretive model nor a purely
    compiling model can provide low latency \emph{and} high throughput, adaptive solutions emerged.  Adaptive systems
    seamlessly transition from interpreted to compiled execution, achieving low latency for short-lived queries and high
    throughput for long-running queries.  However, these adaptive systems pose an immense development effort and require
    expert knowledge in both interpreter and compiler design.

    In this work, we investigate query execution by compilation to \Wasm.  We are able to compile even complex queries
    in less than a millisecond to machine code with near-optimal performance.  By delegating execution of \Wasm to the
    V8~engine, we are able to seamlessly transition from rapidly compiled yet non-optimized code to thoroughly optimized
    code during execution.  Our approach provides both low latency and high throughput, is adaptive out of the box, and
    is straight forward to implement.  The drastically reduced compilation times even enable us to explore generative
    programming of library code, that is fully inlined by construction.  Our experimental evaluation confirms that our
    approach yields competitive and sometimes superior performance.  \done\todo{Should we provide some precise numbers
    from our experimental evaluation?}

\end{abstract}

\maketitle

\pagestyle{\vldbpagestyle}
\begingroup\small\noindent\raggedright\textbf{PVLDB Reference Format:}\\
\vldbauthors. \vldbtitle. PVLDB, \vldbvolume(\vldbissue): \vldbpages, \vldbyear.\\
\href{https://doi.org/\vldbdoi}{doi:\vldbdoi}
\endgroup

\vspace*{8em}

\section{Introduction}
\label{sec:intro}


\begin{figure}[t!]
    \centering
    \includegraphics[scale=.75]{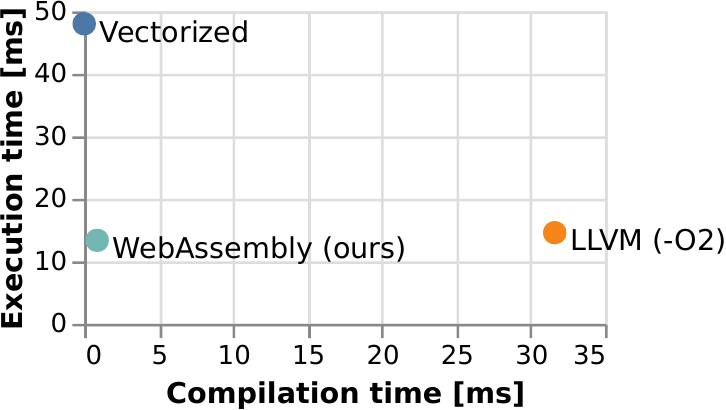}
    \vspace*{-5pt}

    \caption{Design space of query execution engines, based on TPC-H Q6 benchmark results.  The compilation time is the
    time to translate a QEP to executable machine code.  The execution time is the time to execute the machine code and
does not include the compilation time.}
    \label{fig:intro-design_space}\done\todo{Use actual times.}
\end{figure}

\noindent To execute SQL queries, database systems must determine for each query a \emph{query execution plan} (QEP)
that defines how to execute the query.  The QEP is then executed by either interpretation or compilation.  The very
first relational database system, System~R, already compiled QEPs directly to machine code~\cite{chamberlin1981systemr}.
This design decision was not maintained for long and following work dropped compilation in favor of
interpretation~\cite{klonatos2014building}.  The reasoning behind that decision was that compilation directly to machine
code is very difficult to maintain~\cite{menon2017relaxed}.  To target different hardware architectures, the engineers
of System~R had to implement target-specific compilers, which led to a considerable development and maintenance effort.
To remedy this problem, following work used an interpreter to execute queries~\cite{graefe94volcano}.  The induced
overhead of interpretation was dwarfed by the high costs for data accesses in disk-based systems~\cite{boncz2005monet,
neumann2011efficiently, kersten2018everything}.  However, in modern main memory systems data accesses are significantly
faster and the interpretation overhead suddenly takes a large share in query execution costs~\cite{ailamaki1999time,
neumann2011efficiently}.  Therefore, main memory systems must keep any overheads during query execution at a minimum to
achieve peak performance.  This development was the reason for an extensive body of work on query interpretation and
compilation techniques and sparked a seemingly endless debate which of the two approaches to
prefer~\cite{rao2006compiled, krikellas2010generating, neumann2011efficiently, sompolski2011vectorization,
kersten2018everything}.  Recent work proposes an adaptive approach to query execution, where the database system can
seamlessly transition from interpreted to compiled query execution.  This approach requires \emph{both} a query
interpreter \emph{and} a query compiler that must be interoperable, which is achieved by a particular execution mode
named \emph{morsel-wise execution}~\cite{kohn2018adaptive, leis2014morsel}.  Despite the promising results in that work,
we believe that implementing this approach requires expertise in interpreter and compiler design and poses an immense
development effort.

In this work, we present a new approach to query execution by compiling QEPs to \Wasm, ``a low-level assembly-like
language with a compact binary format that runs with near-native performance''~\cite{WasmMDN}.
In~\autoref{fig:intro-design_space}, we sketch the design space of query execution engines, which must balance execution
time versus compilation time.  With our approach we are able to reduce compilation times by more than an order of
magnitude in comparison to state of the art, e.g.\ LLVM-based compilation of QEPs, while at the same time improving
execution times.  Using \Wasm as compilation target for QEPs presents interesting opportunities but also poses new
challenges.

\textbf{Opportunities.}  \Wasm's compact representation allows for fast code generation and efficient caching
of compiled QEPs.  The shipping of \Wasm code in \emph{modules} enables fast adaption and reuse of previously compiled
QEPs.  \Wasm modules are dispatched to a \Wasm engine that takes care of compilation to machine code and further
optimization.  There is a broad selection of open source \Wasm engines, most of which deal with tasks a database
engineer does not want to be concerned with and what traditional compilers do not deliver out of the box.  A few
features we get \say{for free} by compiling to \Wasm and picking a suitable engine are: (1)~two compilation paths, one
using a fast yet non-optimizing compiler and the other using an optimizing compiler, (2)~hot patching of optimized code,
(3)~caching of previously compiled code, and (4)~profile-guided optimization.  Each of these properties can be
beneficial for the execution of QEPs.  So how do we compile QEPs to \Wasm and how can we embed a \Wasm engine in our
database system?

\textbf{Challenges.}  Compilation of QEPs to \Wasm follows text book patterns and is very similar to compilation to
\llvm.  However, \Wasm modules are dispatched to an engine that executes the code in a virtual machine.  The first
challenge is to pass data and control between the database system and \Wasm without throttling performance.  When
embedding a \Wasm engine into a database system, special care must be taken to communicate tables, indexes, and other
data efficiently to the \Wasm code.\done[skip]\todo{A few more words to elaborate the solution, as for the second
challenge.} The second challenge is the fact that \Wasm does not provide a standard library, meaning that data
structures like hash tables, algorithms like sorting, and even basic routines such as \texttt{memcmp} are not available
out of the box.  We could pre-compile a library from another language to \Wasm, e.g.\ \texttt{libc}, and link to that.
However, this approach has two problems.  On the one side, it would prevent inlining of library routines.  On the other
side, generic algorithms and data structures, such as those found in the STL, cannot be translated ahead of time because
\Wasm does not support generic programming.  To circumvent this obstacle, we could compile only the instances of generic
components required by the QEP by providing their type parameters, effectively doing \emph{monomorphisation}.  However,
running a full compilation pipeline from a high-level language to \Wasm contradicts our goal of fast compilation.  We
solve the entire problem of not having a library by doing ad-hoc code generation.  \emph{Every algorithm and data
structure required by a QEP is generated during compilation}.  This is done in such a way, that the concrete types of
generic components, as required in the QEP, are provided to the code generation process, which directly produces the
monomorphic code.  Our approach allows us to rapidly generate code that is already fully inlined and specialized for the
data types used in the QEP.  We are able to achieve performance improvements that, in some cases, can have a tremendous
impact.



\filbreak
\textbf{Contributions.}
\begin{enumerate}
    \item We propose JIT compilation of QEPs to \Wasm, improving both compilation and running time of
        queries.
    \item We embed a suitable \Wasm engine, such that we are able to delegate difficult tasks that require expert
        compiler knowledge to an off-the-shelf system, making our execution engine adaptive out of the box.
    \item We introduce ad-hoc code generation of highly specialized algorithms and data structures and are able to
        outperform traditional approaches that use a pre-compiled library by up to~$4x$.
\end{enumerate}

\textbf{Outline.} This paper is structured as follows.  \autoref{sec:wasm} presents \Wasm and motivates why it is a
suitable compilation target for QEPs.  \autoref{sec:compiling} elaborates compiling QEPs to \Wasm.  In
\autoref{sec:codegen}, we present our ad-hoc generation of specialized library code.  We explain how we execute a
compiled query within a \Wasm engine in \autoref{sec:v8}.  We conduct a comparison to related work in
\autoref{sec:related} and present our experimental evaluation in \autoref{sec:eval}.  \autoref{sec:conclusion} concludes
our work.

\section{\Wasm}
\label{sec:wasm}

To execute queries, we compile QEPs to \Wasm.  This section motivates our choice for \Wasm as compilation target for
QEPs.  In \autoref{sec:wasm-overview} we give the reader an introduction to \Wasm.  We present a small example of \Wasm
in \autoref{sec:wasm-example}.  Finally, in \autoref{sec:wasm-db}, we argue why \Wasm is well-fitted for compiling QEPs.

\subsection{WebAssembly Overview}
\label{sec:wasm-overview}

\Wasm, or short \emph{Wasm}, is a portable binary instruction format~\cite{WasmMDN}.  Among the many high-level goals of
\Wasm, we see three key features that make it the instrument of choice for JIT compiling QEPs.  The first key feature is
that \Wasm is size- and load-time efficient, allowing for fast code generation, fast JIT compilation to machine code,
and resource-friendly caching of already compiled code.  Second, \Wasm can be compiled to execute at native speed and
make use of modern hardware capabilities, e.g.\ SIMD~\cite{V82008}.  The fact that \Wasm is embeddable is the third key
feature for us.  As a bonus, \Wasm can be represented in the \emph{WebAssembly text format} (WAT), a textual
representation for humans.
Now that we introduced \Wasm, let us have a look at a small example and then explore the benefits of using \Wasm as an
intermediate step for compilation of QEPs.

\subsection{\Wasm by Example}
\label{sec:wasm-example}

To give the reader a better understanding of \Wasm, let us have a look at a small example.  We compile the filter
$\sigma_{\text{R.val} < 3.14}$ to \Wasm and explore the compiled code.  \autoref{lst:wasm-example} shows the compiled
code in WAT format\footnote{To be precise, we use the WAST syntax.  This is WAT with symbolic expressions.}, which
borrows its syntax from \textsc{Lisp}, a programming language with fully parenthesized prefix notation.  Due to of lack
of space and the verbosity of WAT, we omit showing boilerplate code for the table scan.  The complete example can be
found in \autoref{apx:wasm-example}.

\begin{lstfloat}
\begin{lstlisting}[language=WebAssembly,mathescape=false]
(; import address of R.val column ;)
(local.set $1 (global.get $R.val))
(; omitted: scan table R ;)
    (if
     (f64.lt
      (f64.load (local.get $1))
      (f64.const 3.14)
     )
     (block $filter.accept
      (; omitted: remainder of the pipeline ;)
     )
    )
    (; advance pointer $1 to next row ;)
    (local.set $1
     (i32.add (local.get $1) (i32.const 4))
    )
\end{lstlisting}
    \caption{\Wasm for the filter $\sigma_{\mathrm{R.val < 3.14}}$ in WAT format.}
    \label{lst:wasm-example}
\end{lstfloat}

In our example, we assume table~\texttt{R} is stored in columnar layout.  Line 2 imports the address of column
\texttt{val} into the local variable~\texttt{\$1}.  This variable is used as a pointer to scan the \texttt{val} column.
Line~3 displays a scan of table~\texttt{R} using a loop construct, which we omitted for brevity.  Line~4 defines a
conditional branch via an \texttt{if}-expression that takes two arguments, first the condition and second the expression
to execute if the condition evaluates to true.  The condition, ranging from line~5 to line~8, defines a 64-bit floating
point comparison.  The first operand of the comparison, found in line~6, is a load instruction from
pointer~\texttt{\$1}.  The second operand is the constant~$3.14$ in line~7.  If the condition evaluates to true, the
block \texttt{\$filter.accept} is executed, where the projection is performed.  Eventually, lines~14 to~16 advance the
pointer~\texttt{\$1} to the next row.

In contrast to the verbose textual format, the bytecode of \Wasm is very compact.  The entire example, including the
omitted parts, is only $219$~consecutive bytes.  The compact encoding allows for fast translation \emph{to} and fast
processing \emph{of} \Wasm and further requires only little memory for caching compiled QEPs.

\subsection{\Wasm in a Database System}
\label{sec:wasm-db}

Database systems must be resilient to different workloads.  Short-running queries should be executed without delay, QEPs
of long-running queries should be compiled and the generated code optimized thoroughly, and for recurring workloads the
generated code should be cached and reused.  In the following, we explain how we fulfill all of these requirements by
compiling QEPs to \Wasm and delegating execution to a suitable \Wasm engine.

As mentioned in the introduction, there is a broad selection of open source \Wasm engines.  To give the reader an
impression, let us name a few prominent examples: Google's~V8~\cite{V82008}, used in NodeJS and Chrome, Mozilla's
SpiderMonkey~\cite{SpiderMonkey}, used in Firefox, Apple's WebKit~\cite{WebKit}, used in Safari, and
Wasmtime~\cite{Wasmtime} and Wasmer~\cite{Wasmer}, both dedicated \Wasm runtimes.  After experimenting with V8 and
SpiderMonkey, we eventually decided to use V8.  Both engines provide similar features and are supported by large
communities.  However, V8 provides a more detailed documentation and insightful examples.  The fact that V8 is written
in \Cpp\ -- the same language that we implement our evaluation system in -- makes embedding V8 easier.

With the following features of V8, we are able to fulfill the requirements mentioned in the beginning of
\autoref{sec:wasm-db}.  V8 provides a two-tier JIT compiler for \Wasm.  The first tier, the \noun{Liftoff} compiler,
achieves low start-up time by generating code as fast as possible at the sacrifice of code quality.  Importantly,
\noun{Liftoff} generates code \emph{in a single pass} over the \Wasm bytecode.  The second tier, the \noun{TurboFan}
compiler, recompiles \emph{hot} code (code that is executed frequently) and performs advanced optimizations.  After
compilation with \noun{TurboFan}, V8 performs \emph{hot patching}:  while the code generated by \noun{Liftoff} is
executing, it is replaced by the optimized code generated by \noun{TurboFan}.  With this procedure, the less performing
code generated by \noun{Liftoff} only runs for short and is gradually replaced by optimized code.  This combination of
two-tier compilation and hot patching of optimized code is a cornerstone of our approach.  Short-running queries are
compiled to \Wasm and delegated to V8, where execution will start without significant delay.  Hence, we are able to
achieve low latencies for short-running queries.  For long-running queries, the optimized code generated by
\noun{TurboFan} eventually takes over execution and the query executes at maximum throughput.  We want to emphasize that
the transition from unoptimized to optimized code is done gradually during execution and handled fully by V8.  By
delegating queries to V8, we provide a query execution engine that is adaptive out of the box.  In contrast to the work
by \ca{kohn2018adaptive}, our approach does not require us to implement this mechanism ourselves.


For recurring queries, we can rely on V8's code caching facilities.  Based on internal heuristics, V8 automatically
performs in-memory caching of the compiled and optimized code produced by \noun{TurboFan}.  When the same \Wasm code is
dispatched multiple times, V8 retrieves the already compiled code from its cache and skips both \noun{Liftoff} and
\noun{TurboFan} compilation.  Caching avoids any potential delay caused by compilation and allows us to immediately
start execution of the optimized code.  In addition to built-in caching, V8 allows for explicit caching by providing
cache data that can be persisted, e.g.\ on disk.

We have presented many benefits of compiling QEPs to \Wasm and delegating execution to V8.  However, whether our
approach is successful depends on the following two questions:  (1)~Can we rapidly compile QEPs to \Wasm? and (2)~Is the
\Wasm code we produce competitive to other systems?  The answer to the first question is a clear `Yes' and elaborated in
\autoref{sec:compiling} and \autoref{sec:codegen}.  To find an answer to the second question, consult our experimental
evaluation in \autoref{sec:eval}.\done[skip]\todo{Rephrase last few sentences; don't artificially build up tension.  May
even leave it out entirely.}

\section{Compiling SQL to \textsc{WebAssembly}}
\label{sec:compiling}

In this section, we elaborate how to compile QEPs of SQL queries to \Wasm.  We follow the approach of
\ca{neumann2011efficiently} in that we dissect a QEP into pipelines, for which we generate code in topological order.
We briefly revisit this approach in \autoref{sec:compiling-pipeline}.  In \autoref{sec:compiling-wasm} we sketch how we
compile algebraic operators to \Wasm.  We motivate our design decisions and provide justifications whenever we deviate
from common practice.


\subsection{Pipeline Model}
\label{sec:compiling-pipeline}

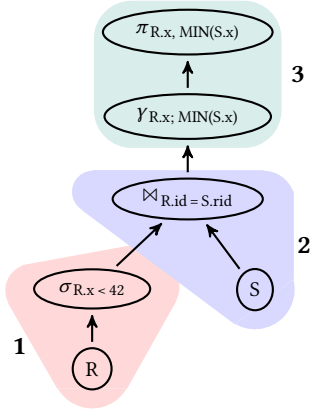
\begin{figure}[t]
    \centering
    \begin{tikzpicture}[node distance=5mm]
        \tikzstyle{op}=[ellipse,thick,draw,inner sep=1mm]

        \node[op] (p) {$\pi_\text{\,R.x, MIN(S.x)}$};
        \node[op,below=of p] (g) {$\gamma_\text{\,R.x; MIN(S.x)}$};
        \node[op,below=of g] (j) {$\Bowtie_\text{\,R.id\,=\,S.rid}$};
        \node[op,below left=8mm and 0mm of j] (s) {$\sigma_\text{\,R.x\,<\,42}$};
        \node[op,below=of s] (R) {R};
        \node[op,below right=8mm and 0mm of j] (S) {S};

        \draw[next] (R) -- (s);
        \draw[next] (s) -- (j);
        \draw[next] (S) -- (j);
        \draw[next] (j) -- (g);
        \draw[next] (g) -- (p);

        \begin{pgfonlayer}{bg}
            \path[fill=red!15]
                (R) ++(-150:15pt) coordinate(R1) arc (-150:-30:15pt) coordinate(R2)
                (s) ++(190:32pt) coordinate(s1) arc (210:110:10pt) coordinate(s2)
                (s) ++(-8:32pt) coordinate(s3) arc (-30:45:10pt) coordinate(s4)
                (s) ++(30:27pt) coordinate(p0_1) arc (45:100:10pt) coordinate(p0_2);
            \path[draw=red!15,fill=red!15]
                (R2) -- (s3) --
                (s4) -- (p0_1) --
                (p0_2) -- (s2) --
                (s1) -- (R1);

            \path[fill=blue!15]
                (S) ++(-120:15pt) coordinate(S1) arc (-120:15:15pt) coordinate(S2)
                (j) ++(4:40pt) coordinate(j1) arc (20:90:12pt) coordinate(j2)
                (j) ++(186:40pt) coordinate(j3) arc (240:90:8pt) coordinate(j4);
            \path[draw=blue!15,fill=blue!15]
                (S2) -- (j1) --
                (j2) -- (j4) --
                (j3) -- (S1);

            \node[fill=PineGreen!15,thick,fit=(g) (p),rounded corners=5mm] (p3) {};
        \end{pgfonlayer}

        \node[left=of R,yshift=3mm] {\large\bfseries 1};
        \node[right=2mm of S,yshift=6mm] {\large\bfseries 2};
        \node[right=0mm of p3] {\large\bfseries 3};
    \end{tikzpicture}
    \caption{A QEP for the query in \autoref{lst:compiling-pipeline-sql}, containing three pipelines enumerated in
        topological order.
    }
    \label{fig:compiling-pipeline}
\end{figure}

A QEP is -- in its most essential form -- a tree with tables or indexes at the leaves and algebraic operators at the
inner nodes.\footnote{The authors are aware that a QEP need not strictly be a tree and in some situations a
representation as directed acyclic graph is desirable~\cite{neumann2015unnesting}.}  \autoref{fig:compiling-pipeline}
shows a QEP for the query in \autoref{lst:compiling-pipeline-sql}.  The edges between nodes of the tree point in the
direction of data flow.

\begin{lstfloat}[t]
    \begin{lstlisting}[language=sql]
        SELECT R.x, MIN(S.x)
        FROM R, S
        WHERE R.x < 42 AND R.id = S.rid
        GROUP BY R.x;
    \end{lstlisting}
    \caption{Example query to demonstrate the pipeline model.}
    \label{lst:compiling-pipeline-sql}
\end{lstfloat}

The tree structure of a QEP can be dissected into \emph{pipelines}.  A pipeline is a linear sequence of operators that
does not require materialization of tuples.  To identify the pipelines of a QEP, we hence must identify all operators
that require materialization, named \emph{pipeline breakers}.  The most common pipeline breakers are grouping, join, and
sorting; table scan, index seek, selection, and projection are not pipeline breakers.

In \autoref{fig:compiling-pipeline}, we have colored and enumerated the three pipelines of the QEP.  Pipeline~1 scans
table~\lstinline{R}, selects all tuples where \lstinline{R.x < 42}, and inserts all qualifying tuples into a hash table
for the join.  Pipeline~2 scans table~\lstinline{S} and probes all tuples against the hash table constructed by
pipeline~1.  Every pair of tuples from \lstinline{R} and \lstinline{S} that satisfies the condition
\lstinline{R.id = S.rid} is joined and inserted into another hash table where groups of \lstinline{R.x} are formed.
Pipeline~3 iterates over these groups and performs the final projection.

After dissecting the QEP into pipelines, each pipeline is compiled separately.  However, we must order the pipelines
such that all data dependencies of the QEP are satisfied.  For example, pipeline~3 iterates over all groups produced by
grouping.  Hence, pipeline~2 that forms those groups must be executed \emph{before} pipeline~3.  By topologically
sorting the pipelines we compute an order that satisfies all data dependencies in the QEP.

The pipeline model allows us to dissect a QEP into linear sequences of operators that process tuples without need for
intermediate materialization.  The pipeline model does \emph{not} dictate whether to push or pull tuples, whether to
process tuples one at a time or in bulk, or whether to execute the QEP by compilation or interpretation.  In this work,
we compile the pipelines of a QEP such that a single tuple is pushed at a time through the entire pipeline until it is
materialized in memory.

\subsection{Compiling to \Wasm}
\label{sec:compiling-wasm}

To compile QEPs to \Wasm, we follow the approach of \ca{neumann2011efficiently}.  Generating \Wasm code is very similar
to generating \llvm code, for example.  In the following, we briefly sketch how to compile the most common operators of
a QEP to \Wasm.  We elaborate whenever our approach deviates from regular compilation.

\noindent \textbf{Table scan, index seek, and pipeline breakers} --- The start of a pipeline -- which is either a table
scan, an index seek, or a pipeline breaker -- is translated to a loop construct.  For a table scan, we emit code to
access all tuples of the respective table.  For an index seek, we emit code to iterate over all qualifying entries in
the respective index.  For a pipeline breaker, e.g.\ grouping, we emit code to iterate over all materialized tuples,
e.g.\ groups.  The remainder of the pipeline is compiled into the loop's body.

\noindent \textbf{Selection} --- A selection is compiled to a conditional branch.  The condition is compiled without
short-circuit semantics, i.e.\ both sides of a logical conjunction or disjunction are evaluated.  It is debatable
whether to prefer short-circuit evaluation.  For \say{simple} predicates, short-circuit evaluation is likely a bad
choice.  It introduces a conditional branch that unnecessarily stresses branch
prediction~\cite{sompolski2011vectorization}.  It may further lead to a conditional load from memory, which may
negatively impact prefetching~\cite{kersten2018everything}.  For \say{complex} predicates, short-circuit evaluation
likely pays off.  The additional conditional branch can bypass costly evaluation of the right hand side of a logical
conjunction or disjunction~\cite{ross2002conjunctive}.  This judgement is reflected in many compilers, e.g.\
\noun{Clang} and \noun{GCC}, which remove the short-circuit evaluation from C~code during compilation if the predicates
are presumably cheap to compute (and without side effects).  This transformation is a part of
\emph{if-conversion}~\cite{allen1983conversion}.  We implement a very simplistic estimation, treating all comparisons of
numeric and boolean types as \say{cheap} and comparisons of character sequences as costly relative to the sequence
length.

\noindent \textbf{Projection} --- The projection of an attribute or aggregate does not require an explicit operation.
The code necessary to access the attribute's or aggregate's value has already been generated when compiling the
beginning of the pipeline.  To compile the projection of an expression, we compile the expression and assign the result
to a fresh local variable.  In contrast to interpretation, projecting attributes \emph{away} is performed implicitly and
requires no further code.  Because the attribute that is projected away is not used further up in the QEP, no code using
the attribute is generated.  The register or local variable holding the attribute's value is automatically reclaimed
during compilation to machine code~\cite{boissinot2008fast}.

\noindent \textbf{Hash-based Grouping \& Aggregation} --- Hash-based grouping is a pipeline breaker.  The incoming
pipeline to the grouping operator assembles the groups in a hash table and immediately updates the group's aggregates.
The pipeline starting at the grouping operator iterates over all assembled groups as explained above.

An important distinction between this and previous work is how inserts and updates to the hash table are performed.
Previous work~-- including both interpretation- and compilation-based execution~-- relies on the existence of a
pre-compiled library that provides a hash table implementation.  This means, calls to the hash table must use an
interface that is agnostic to the type of data being inserted.  This imposes an artificial constraint on the library.
Because the type of a hash table entry is not known at the time when the library is compiled, certain hash table designs
cannot be implemented effectively.  To lookup a key we first must compute the key's hash.  This computation can be
performed \emph{outside} the library and the computed hash value can be passed through the hash table's interface, as
done by \ca{neumann2011efficiently}.  However, we must search the collision list for an entry with the same key.
Because the hash table's interface is type-agnostic, the hash table does not know how to compare two keys.  Hence, we
must provide a comparison function to the hash table's lookup function.  This means, for every comparison of two keys, a
comparison function must be called.  (To lookup $n$ keys, at least $n$ such calls are necessary!)  The situation gets
worse if the hash table must be able to grow dynamically.  To grow a hash table, all elements of the table must be
rehashed.  Again, because the hash table is type-agnostic, it does not know how to hash the elements.  Hence, a hash
function must be provided in addition to a comparison function or the computed hash values must be stored in the hash
table.  Another downside of using a pre-compiled library is that calls to the library cannot be inlined.  Hence, every
access to the hash table requires a separate function call.

We resolve these issues by generating and JIT compiling the code for the hash table during compilation of the QEP.
Although this sounds very expensive and prohibitive, we show in \autoref{sec:eval} that generating and compiling \Wasm
is so fast that it becomes feasible at running time.  We explain the generation of library code in detail in
\autoref{sec:codegen}.

\noindent \textbf{Simple Hash Join} --- A simple hash join is a pipeline breaker for one of its inputs.  The incoming
pipeline, by convention the left subtree of the join, inserts tuples into a hash table.  The pipeline of the join probes
its tuples against that hash table to find all join partners.  The same distinction between this and previous work as
for Grouping \& Aggregation applies here.  To avoid artificial constraints on hash table design and to avoid issuing a
function call per access to a hash table, we generate and JIT compile the required hash table code during compilation of
the QEP.  This approach is elaborated in \autoref{sec:codegen}.

\noindent \textbf{Sorting} --- Sorting is a pipeline breaker and very similar to Grouping \& Aggregation.  Before the
sorting operator can produce any results, all tuples of the incoming pipeline must be produced and materialized.  After
the incoming pipeline has been processed entirely, the sorting operator can output tuples in the specified order.

We implement the sorting operator by collecting all tuples from the incoming pipeline in an array and sorting the array
with \noun{Quicksort}.  The way we integrate sorting into the compiled QEP is an important distinction between this and
previous work.  In previous work that performs compilation, a sorting algorithm already exists as part of a pre-compiled
library that is invoked to sort the array.  The interface to this sorting algorithm is type-agnostic, i.e.\ the sorting
algorithm does not know what it is sorting.  In order to compare and move elements in the array, additional information
must be provided when invoking the sorting algorithm.  For comparison-based sorting, the size of an element in the array
and a function that computes the order of two elements must be provided.  This is very well exemplified by
\lstinline{qsort} from \texttt{libc}.  This design leads to two severe performance issues.  First, because the size of
the elements to sort is not known when the library code is compiled, a generic routine such as \lstinline{memcpy} must
be used to move elements in the array.  This may result in suboptimal code to move elements or even an additional
function call per move.  Additionally, values cannot be passed through registers and must always be read from and
written to memory, obstructing optimization by the compiler.  Second, to compute the order of two elements an external
function must be invoked.  This means, for every comparison of two elements the sorting algorithm must issue a separate
function call.  (To sort $n$ elements, at least $\Theta \left(n\,\textit{log}\,n\right)$ such calls are necessary!)

When the QEP is being interpreted, e.g.\ in the vectorized execution model, similar problems emerge.  Although tuples
need not be moved if an additional array of indices is used, the sorting algorithm must delegate the comparison of two
tuples to the interpreter, where the predicate to oder by is dissected into atomic terms that are evaluated separately.
This leads to significant interpretation overhead at the core of the sorting algorithm.

We resolve the aforementioned issues by generating and JIT compiling the library code during compilation of the QEP.
Our generated sorting algorithm is precisely tuned to the elements to sort and the order to sort them by.  In
particular, the comparison of two elements is fully inlined into the sorting algorithm.  We explain this approach in
detail in \autoref{sec:codegen}.

\section{Library Code Generation}
\label{sec:codegen}

In \autoref{sec:intro} and \autoref{sec:compiling-wasm} we already motivated our decision to generate specialized
library code just in time during compilation of a QEP.  In this section, we present our process of ad-hoc library code
generation along the example of generating specialized \noun{Quicksort}.  We begin with partitioning and inlined
comparison of elements before we explain how we generate \noun{Quicksort}.

\subsection{Conceptual Comparison}

Before diving into the code generation example, let us reconsider our approach on a conceptual level and compare it with
alternatives.  A problem that is inherent in all query execution engines is that their supported operations must be
polymorphic.  Joins, grouping, sorting, etc.\ must be applicable to attributes of any type and size.  We aim to provide
this polymorphism at query compilation time by generating specialized library code.  To understand how other systems
solve this task, let us have a look at state-of-the-art solutions.

\begin{lstfloat}[t]
    \begin{lstlisting}[language=C++]
    /* Create a fresh vector with indices
     * from 0 to VECTOR_SIZE - 1. */
    sel0 = create_selection_vector(VECTOR_SIZE);
    /* Evaluate LHS of conjunction. */
    sel1 = cmp_lt_i32_imm(sel0, vec_R_x, 42);
    /* Evaluate RHS of conjunction. */
    sel2 = cmp_gt_i64_imm(sel1, vec_R_y, 13);
    \end{lstlisting}
    \caption{Demonstration of the vectorized processing model.  The condition
    \lstinline[language=sql]{R.x < 42 AND R.y > 13} is computed by successively refining a selection vector.}
    \label{lst:codegen-vectorized}
\end{lstfloat}

\textbf{Vectorization.}  In the vectorized processing model, operations are specialized and pre-compiled for the
different types of vectors.  In \autoref{lst:codegen-vectorized}, we provide an example for the evaluation of a
selection with a conjunctive predicate.  The initial selection vector \lstinline{sel0} is successively refined by calls
to vectorized comparison functions \lstinline{cmp_*} and eventually \lstinline{sel2} contains all indices where the
selection predicate is satisfied.  A vectorized query interpreter executes a QEP by calling these vectorized functions
and managing the data flow between function calls.  To achieve short-circuit evaluation of the condition, the selection
vector \lstinline{sel1} is passed to the second comparison, such that the right-hand side of the conjunctive predicate
is only evaluated for elements that also satisfy the left-hand side.  In a compiling setting, short-circuit evaluation
is usually implemented as a conditional branch.  In the vectorized processing model, that control flow is converted to
data flow.  Conditional control flow can benefit from branch prediction, which works well in either case when the
selectivity is very high or very low.  However, when the control flow is converted to data flow, the benefit on low
selectivities is lost~\cite{sompolski2011vectorization, pirk2016voodoo, kersten2018everything}.  This is very well
exemplified by our example in \autoref{lst:codegen-vectorized}.  Assume that the left hand side of the condition is
barely selective.  Although the outcome of evaluating the left hand side can be well predicted, evaluation of the right
hand side in line~7 can only start once the comparison in line~5 completes.  Hence, this design completely eliminates
the processors ability to predict the outcome of evaluating the left-hand side and executing the right-hand side
unconditionally and out of order, as opposed to how it would be in a compiling setting.  Another drawback of the
vectorized processing model is the fact that operations must be specialized and compiled ahead of time.  It is
infeasible to provide vectorized operations for arbitrary expressions, as there are infinitely many.  Therefore, the
interpreter dissects expressions into atomic terms for which a finite set of vectorized operations is pre-compiled.  For
our example in \autoref{lst:codegen-vectorized}, this means that the interpreter must \emph{always} evaluate one side of
the conjunction after the other and cannot evaluate both sides at once.

\begin{lstfloat}[t]
    \begin{lstlisting}[language=C++]
    /* Initialize hash table. */
    HT *ht = lib_HT_create();
    /* Iterate over all rows of table R. */
    for (auto row : tbl_R) {
      /* Evaluate selection predicate. */
      if (row.x < 42 and row.y > 13) {
        /* Compute hash of R.id. */
        auto hash = ... row.id ...;
        /* Insert into hash table. */
        char *ptr=lib_HT_insert(ht, hash, 8);
        *(int*) ptr    = row.id; // key
        *(int*)(ptr+4) = row.x;  // value
      }
    }
    \end{lstlisting}
    \caption{Demonstration of a compilation-based processing model with calls to a pre-compiled library.  Every
      insertion into the hash table requires a separate function call.}
  \label{lst:codegen-hash}
\end{lstfloat}

\textbf{Linking with pre-compiled library.}  In a compilation-based processing model, e.g.\ \noun{HyPer}, every
operation in the QEP is compiled to a code fragment.  The produced code is specific to the types of the operation's
operands.  Arbitrarily complex expressions are compiled directly rather than taking a detour through a pre-compiled
library by function calls.  Thereby, the compiler can choose to implement short-circuit evaluation by conditional
control flow.

The biggest drawback of compiling QEPs is the time spent compiling.  While direct compilation to machine code could be
done rapidly, the produced code would certainly be of poor quality.  Therefore, compilation-based systems employ
compiler frameworks like \noun{LLVM} to perform optimizations on the code.  While these optimizations can greatly
improve the performance of the code, they require costly analysis and transformation.  Hence, compilation of queries can
easily take more than a hundred milliseconds~\cite{kohn2018adaptive}.

To reduce the amount of code that must be compiled, recurring routines such as hash table lookups or sorting are
pre-compiled and shipped in a library.  During compilation of a QEP, when an operation can be delegated to a
pre-compiled routine, the compiler simply produces a respective function call to the library.  This is a trade-off
between compilation time and running time and the biggest drawback of this approach.  Function calls to a pre-compiled
library prevent inlining and obstruct further optimization, and thereby potentially lead to sub-optimal performance.  We
demonstrate this in \autoref{lst:codegen-hash}, where every insertion into a hash table requires a separate function
call.  The library code for probing the hash table can be compiled and optimized thoroughly ahead of time.  Because the
size of a hash table entry is not known when the library is compiled, the size must be provided at running time when
inserting an entry.  In the example, the hash table must allocate 8~bytes per entry to store \lstinline|R.id| and
\lstinline|R.x| and it is the task of the caller to assign those values to the entry.

\textbf{Full compilation.}  In this approach, code for the entire QEP with all required algorithms and data
structures is generated and compiled just in time.  By generating the code just in time, it is possible to produce
highly specialized code, target particular hardware features, and enable holistic optimization.  One example for full
compilation is template expansion, as done in the \textsc{Hique} system~\cite{krikellas2010generating}.  \textsc{Hique}
provides a set of generic algorithms and data structures that are instantiated and compiled to implement the QEP.
Another example is code generation via \emph{staging}, as done in \textsc{LegoBase}.  Here, metaprogramming is used to
write a query engine in Scala~LMS, that when partially evaluated on an input QEP outputs specialized C~code that
implements the query~\cite{klonatos2014building}.  While full compilation can achieve the highest possible throughput,
both \textsc{Hique} and \textsc{LegoBase} take considerable compilation time with hundreds of milliseconds for single
TPC-H queries.

\subsection{Our approach: JIT code generation}

\begin{figure}[t]
    \centering
    \begin{tikzpicture}[node distance=0mm,every node/.style={align=center}]
        \coordinate (start);
        \node[block,right=8mm of start] (compiler) {\Wasm\\Compiler};
        \node[block,above=14mm of compiler] (codegen) {\Wasm\\Codegen};
        \node[block,right=16mm of compiler] (platform) {\Wasm\\Platform};
        \node[block,above=14mm of platform] (v8) {V8};
        \coordinate[right=13mm of platform] (end);

        \draw[next] (start) to node[above,ann] {QEP} (compiler);

        \draw[next,transform canvas={xshift=-2mm}] (compiler) to node[left,ann] {type information} (codegen);
        \draw[next,transform canvas={xshift=2mm}] (codegen) to node[right,ann,align=left,yshift=4mm]
            {specialized algorithms\\and data structures} (compiler);

        \draw[next] (compiler) to node[above,ann,align=center] {\Wasm\\module} (platform);
        \draw[next,transform canvas={xshift=-2mm},dashed] (platform) to node[left,ann,align=right,yshift=-2mm]
            {\Wasm module\\and data} (v8);

        \draw[next,dashed,out=30,in=-30,looseness=5] (v8) to node[right,ann,align=center]
            {compile to x86\\\&\ execute} (v8);
        \draw[next,dashed,transform canvas={xshift=2mm}] (v8) to node[right,ann,align=left] {Module memory} (platform);

        \draw[next] (platform) to node[above,ann,align=center] {result set} (end);
    \end{tikzpicture}

    \caption{Architecture of our query execution engine.  QEPs are compiled to \Wasm modules.  Specialized code is
        generated on demand just in time for algorithms and data structures required by the QEP.  \Wasm modules are
        dispatched to the \Wasm platform, that passes required data and delegates execution to Google's~V8.  V8~compiles
        and executes the modules, producing results in the module's memory, where they are extracted by the platform.}
    \label{fig:compiling-arch}
\end{figure}
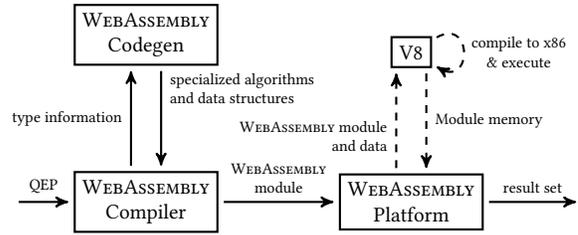

Full compilation is very similar to our approach of generating required library routines just in time and JIT compiling
the QEP.  The key distinction is how the generation of code is performed.  Previous work generates code in a high-level
language.  This code must then go through parsing and semantic analysis before it is translated to a lower level
intermediate representation where optimizations are performed before executable machine code is produced.  Going through
the entire compiler machinery takes a lot of time.  Our approach, depicted in \autoref{fig:compiling-arch}, bypasses
most of these steps.  We generate specialized algorithms and data structures directly in \Wasm.  By picking a suitable
\Wasm engine, e.g.\ V8, we get all the benefits described in \autoref{sec:wasm-db}.  Our approach is able to produce
highly specialized algorithms and data structures and enables holistic optimization without the drawback of long code
generation and compilation times.

\subsection{Code Generation by Example}
\label{sec:codegen-example}

To provide the reader with a better understanding of how we generate library code just in time, let us exemplify our
code generation along the example of \noun{Quicksort}.  We build the example bottom up, beginning with code generation
for partitioning and the comparison of two elements before we explain code generation of the recursive \noun{Quicksort}
algorithm.

\begin{lstfloat}[t]
    \caption{Pseudo code for the generation of specialized code that implements \textsc{Hoare's partitioning}.}
    \label{algo:hoare}
    \begin{algorithmic}[1]
    \Function{Partition}{\textit{order}, \textit{begin}, \textit{end}, \textit{pivot}}
    \State $l \leftarrow $ \Call{NewVar}{\null}
    \State $r \leftarrow $ \Call{NewVar}{\null}
    \State \Call{Emit}{$l \leftarrow \textit{begin}$}
    \State \Call{Emit}{$r \leftarrow \textit{end}$}
    \State \Call{Emit}{\textbf{while} $l < r$}
    \State \Call{EmitSwap}{$l$, $r-1$}
    \State $cl \leftarrow$ \Call{EmitCompare}{\textit{order}, $l$, \textit{pivot}}
    \State $cr \leftarrow$ \Call{EmitCompare}{\textit{order}, \textit{pivot}, $r-1$}
    \State \Call{Emit}{$l \leftarrow l + cl$}
    \State \Call{Emit}{$r \leftarrow r - cr$}
    \State \Call{Emit}{\textbf{end while}}
    \State \textbf{return} $l$
    \EndFunction
    \end{algorithmic}
\end{lstfloat}

\textbf{\noun{Hoare's partitioning} scheme.}
\noun{Hoare's partitioning} scheme creates two partitions from a sequence of elements based on a boolean predicate such
that all elements in the first partition do not satisfy the predicate and all elements in the second partition satisfy
the predicate.  We apply \noun{Hoare's partitioning} in our generated \noun{Quicksort} algorithm, that in turn is used
to implement sorting of tuples.  In our setting, the sequence of tuples to partition is a consecutive array.  The
predicate to partition the array is derived from a list of expressions to order by.  For example, the clause
\lstinline[language=sql]|ORDER BY R.x + R.y, R.z| results in the list $\left[R.x + R.y, R.z\right]$.

We provide pseudo code for the generation of specialized partitioning code in \autoref{algo:hoare}.  The function
\textsc{Partition} takes four parameters: the \textit{order} is a list of expressions to order by, \textit{begin} and
\textit{end} are variables holding the address of the first respectively one after the last tuple in the array to
partition, and \textit{pivot} is a variable holding the address of the pivot to partition by.  The pivot must not be in
the range [\textit{begin}, \textit{end}).  First, the algorithm copies the values of \textit{begin} and \textit{end} by
introducing fresh variables $l$ and $r$ in lines~2 and~3 and then emitting code that assigns the value of \textit{begin}
to $l$ in line~4 and the value of \textit{end} to $r$ in line~5.  Next, in line~6, a loop header with the condition $l <
r$ is emitted.  The code emitted thereafter forms the loop body.  In line~7, \textsc{EmitSwap} is called to emit code
that swaps the tuples at the addresses $l$ and $r-1$.  Note that this is a function call \emph{during code generation}.
The call will emit code directly into the loop body, as if inlined by an optimizing compiler, and there will be no
function call during execution of the generated code.  In lines~8 and~9, \textsc{EmitCompare} is called to emit code
that compares the tuples at addresses $l$ and $r-1$ to the tuple at address \textit{pivot} according to the order
specified by \textit{order}.  Each call returns a fresh boolean variable that holds the outcome of the comparison.  Just
like \textsc{EmitSwap}, calls to \textsc{EmitCompare} emit code directly into the loop body without the need for a
function call in the generated code.  The value of variable $cl$ will be true if the tuple at address $l$ compares less
than the tuple at address \textit{pivot} w.r.t.\ the specified \textit{order}.  Line~10 emits code that advances $l$ to
the next tuple if $cl$ is true, otherwise $l$ is not changed.  Similarly, line~11 emits code to advance $r$ to the
previous tuple if $cr$ is true.  This is a means of implementing branch-free partitioning.  In line~12, the loop body
for the loop emitted in line~6 is finished.  Eventually, \textsc{Partition} returns the variable $l$, which will point
to the beginning of the second partition once the loop of line~6 terminates.

\begin{lstfloat}[t]
    \caption{Pseudo code for the generation of code that compares two elements based on a specified order.}
    \label{algo:compare}
    \begin{algorithmic}[1]
    \Function{EmitCompare}{\textit{order}, $l$, $r$}
    \State $v \leftarrow \Call{NewVar}{\null}$
    \State \Call{Emit}{$v \leftarrow 0$}
    \For{\textbf{each} \textit{expr} \textbf{in} \textit{order}}
        \State $vl \leftarrow$ \Call{Compile}{\textit{expr}, $l$}
        \State $vr \leftarrow$ \Call{Compile}{\textit{expr}, $r$}
        \Switch{$\textit{type}(\textit{expr})$}
        \Case{int}
        \State \Call{Emit}{$lt \leftarrow vl <_\text{int} vr$}
        \State \Call{Emit}{$gt \leftarrow vl >_\text{int} vr$}
        \EndCase
        \Case{float}
        \State \Call{Emit}{$lt \leftarrow vl <_\text{float} vt$}
        \State \Call{Emit}{$gt \leftarrow vl >_\text{float} vt$}
        \EndCase
        \State ... \Comment{cases for remaining types}
        \EndSwitch
        \State \Call{Emit}{$v \leftarrow 2 \cdot v + gt - lt$}
    \EndFor
    \State $c \leftarrow$ \Call{NewVar}{\null}
    \State \Call{Emit}{$c \leftarrow v < 0$}
    \State \textbf{return} $c$
    \EndFunction
    \end{algorithmic}
\end{lstfloat}

The code presented in \autoref{algo:hoare} looks almost like a regular implementation of partitioning.  However, the
function \emph{emits} code that will perform partitioning.  An important part of partitioning, that we skipped in
\autoref{algo:hoare}, is how the code to compare two tuples based on a given order is generated.  Therefore, we also
provide pseudo code for \textsc{EmitCompare} in \autoref{algo:compare}.

First, \textsc{EmitCompare} creates a fresh variable $v$ in line~2 and initializes it to $0$ in line~$3$.  Then, in
line~4, the function iterates over all expressions in \textit{order}.  The call to \textsc{Compile} in line~5 emits code
to evaluate \textit{expr} on the tuple pointed to by $l$ and returns a fresh variable holding the value of the
expression.  Analogously, line~6 evaluates \textit{expr} on the tuple pointed to by $r$.  Next, type-specific code to
compare the values $vl$ and $vr$ of the evaluated expression is emitted.  Because the particular code to emit depends on
the type of \textit{expr}, line~7 performs a case distinction on the type.  This case distinction is performed
\emph{during code generation} and the generated code will only contain the emitted, type-specific code.  In case the
expression evaluates to an \texttt{int}, lines~9 and~10 emit code to perform an integer comparison of $vl$ and $vr$.
The cases for other types are analogous.  After emitting type-specific code for the comparison of $vl$ and $vr$, line~16
emits code to update $v$ based on the outcome of the comparison.  After generating code to evaluate all expressions in
\textit{order} and updating $v$ accordingly, lines~18 and~19 introduce a fresh boolean variable $c$ that will be set to
$v < 0$, which evaluates to true if the tuple at $l$ is strictly smaller than the tuple at $r$, and false otherwise.

\begin{lstfloat}[t]
    \caption{Generated partitioning code for the order $[R.x + R.y, R.z]$.}
    \label{algo:partition}
    \begin{algorithmic}[1]
    \Require{b, e, p}
    \Ensure{$p_l$}
    \State \textbf{var} $p_l \leftarrow$ b \Comment{Initialize pointers to the first and}
    \State \textbf{var} $p_r \leftarrow$ e \Comment{one after the last tuple, respectively.}
    \While{$p_l < p_r$}
    \State \textcolor{AlgoComment}{$\triangleright$ \Call{EmitSwap}{$p_l$, $p_r-1$}}
    \State \textbf{var} $v_\textit{tmp} \leftarrow *p_l$ \Comment{Use temporary variable}
    \State $*p_l \leftarrow *(p_r-1)$ \Comment{to swap tuples}
    \State $*(p_r-1) \leftarrow v_\textit{tmp}$ \Comment{at $p_l$ and $p_r-1$.}
    \State \textcolor{AlgoComment}{$\triangleright$ \Call{EmitCompare}{{\footnotesize$[R.x + R.y, R.z]$}, $p_l$, p}}
    \State \textbf{var} $v_\textit{l,pivot} \leftarrow 0$
    \State \textbf{var} $v_l \leftarrow p_l.x +_\texttt{int} p_l.y$ \Comment{{\footnotesize \Call{Compile}{$R.x + R.y$, $p_l$}}}
    \State \textbf{var} $v_\textit{pivot} \leftarrow \text{p}.x +_\texttt{int} \text{p}.y$ \Comment{{\footnotesize \Call{Compile}{$R.x+R.y$, {\normalfont p}}}}
    \State \textbf{var} $v_\textit{lt} \leftarrow v_l <_\texttt{int} v_\textit{pivot}$
    \State \textbf{var} $v_\textit{gt} \leftarrow v_l >_\texttt{int} v_\textit{pivot}$
    \State $v_\textit{l,pivot} \leftarrow 2 \cdot v_\textit{l,pivot} + v_\textit{gt} - v_\textit{lt}$
    \State \textbf{var} $v_l \leftarrow p_l.z$ \Comment{{\footnotesize \Call{Compile}{$R.z$, $p_l$}}}
    \State \textbf{var} $v_\textit{pivot} \leftarrow \text{p}.z$ \Comment{{\footnotesize \Call{Compile}{$R.z$, {\normalfont p}}}}
    \State \textbf{var} $v_\textit{lt} \leftarrow v_l <_\texttt{int} v_\textit{pivot}$
    \State \textbf{var} $v_\textit{gt} \leftarrow v_l >_\texttt{int} v_\textit{pivot}$
    \State $v_\textit{l,pivot} \leftarrow 2 \cdot v_\textit{l,pivot} + v_\textit{gt} - v_\textit{lt}$
    \State \textbf{var} $v_\textit{cl} \leftarrow v_\textit{l,pivot} < 0$
    \State \textcolor{AlgoComment}{$\triangleright$ \Call{EmitCompare}{{\footnotesize$[R.x + R.y, R.z]$}, p, $p_r-1$}}
    \State \dots \Comment{Code omitted for brevity.}
    \State \textbf{var} $v_\textit{cr} \leftarrow v_\textit{pivot,r} < 0$
    \State $p_l \leftarrow p_l + v_\textit{cl}$ \Comment{Advance left cursor.}
    \State $p_r \leftarrow p_r - v_\textit{cr}$ \Comment{Advance right cursor.}
    \EndWhile
    \end{algorithmic}
\end{lstfloat}

To put it all together, let us exercise an example.  We invoke \textsc{Partition} with the \textit{order} \mbox{$[R.x +
R.y, R.z]$}, \textit{begin} `b', \textit{end} `e', and \textit{pivot} `p'.  The generated code is given in
\autoref{algo:partition}.  Initially, in lines~1 and~2, the addresses of the first and one after the last tuple are
stored in fresh variables.  Then the loop in line~3 repeats as long as pointer~$p_l$ points to an address smaller
than~$p_r$.  Lines~5 to~7 show the code produced by \textsc{EmitSwap}, that swaps two tuples using a temporary variable.
In lines~9 to~20, the tuple at $p_l$ is compared to the pivot according to the specified order.  Variable
$v_\textit{cl}$ is true if the tuple at $p_l$ compares less than the pivot, false otherwise.  Analogously, the tuple at
$p_r - 1$ is compared to the pivot.  To keep the example short and because the code is very similar, we omit this code
and only show a place holder in line~22.  At the end of the loop, in lines~24 and~25, the pointers $p_l$ and $p_r$ are
advanced depending on the outcome of the comparisons.

The generated code will partition the range \mbox{[b, e)} such that the first partition contains only tuples that
compare less than~p and the second partition contains only tuples greater than or equal to~p, w.r.t.\ the specified
order.  Note that the generated code is not a function.  Instead, this code can be generated into a function where
partitioning is needed.  Hence, the entire code for partitioning will always be fully inlined and specialized for the
order to partition by.

\textbf{\noun{Quicksort}.}  \noun{Quicksort} sorts its input sequence by recursive partitioning.  In our implementation
of \noun{Quicksort}, we compute the pivot to partition by as a \emph{median-of-three}.  With our code generation for
partitioning at hand, generating \noun{Quicksort} is relatively simple.  We provide pseudo code in
\autoref{algo:quicksort}.  Line~2 defines a new function \texttt{qsort}, line~3 emits a loop that repeats as long as
there are more than two elements in the range from \textit{begin} to \textit{end}.  Inside this loop, lines~4 to~7 emit
code to compute the median of three and bring the median to the front of the sequence to sort.  Line~8 emits the code to
partition the sequence $\textit{begin} + 1$ to \textit{end} using as pivot the median of three.  After partitioning, the
median must be swapped back into the partitioned sequence, which is done by line~9.  Line~10 checks whether to recurse
into the right partition.  Line~11 emits a recursive call to sort the right partition with \texttt{qsort}.  Afterwards,
in line~13, code is emitted to update \textit{end} to the end of the left partition.  Once the loop emitted by line~3
exits, we must handle the special case that $\textit{end} - \textit{begin} = 2$, omitted here for brevity.

\begin{lstfloat}[t]
    \caption{Pseudo code to generate specialized \noun{Quicksort}.}
    \label{algo:quicksort}
    \begin{algorithmic}[1]
    \Function{Quicksort}{\textit{order}}
    \State \Call{Emit}{\textbf{function} qsort(\textit{begin}, \textit{end})}
    \State \Call{Emit}{\textbf{while} $\textit{end} - \textit{begin} > 2$}
    \State \textit{mid} $\leftarrow$ \Call{NewVar}{\null}
    \State \Call{Emit}{$\textit{mid} \leftarrow \textit{begin} + (\textit{end} - \textit{begin}) / 2$}
    \State $m \leftarrow$ \Call{EmitMedianOf3}{\textit{begin}, \textit{mid}, $\textit{end}-1$}
    \State \Call{EmitSwap}{\textit{begin}, $m$}
    \State $\textit{mid} \leftarrow$ \Call{Partition}{\textit{order}, $\textit{begin}+1$, \textit{end}, \textit{begin}}
    \State \Call{EmitSwap}{\textit{begin}, $\textit{mid}-1$}
    \State \Call{Emit}{\textbf{if} $\textit{end} - \textit{mid} \ge 2$}
    \State \Call{Emit}{qsort(\textit{mid}, \textit{end})}
    \State \Call{Emit}{\textbf{end if}}
    \State \Call{Emit}{$\textit{end} \leftarrow \textit{mid}-1$}
    \State \Call{Emit}{\textbf{end while}}
    \State \textcolor{AlgoComment}{$\triangleright$ handle case where $\textit{end} - \textit{begin} = 2$}
    \State \Call{Emit}{\textbf{end function}}
    \EndFunction
    \end{algorithmic}
\end{lstfloat}

We can see that by executing our \noun{Quicksort} code generation, we obtain a specialized, fully inlined \texttt{qsort}
function that can be called to sort a sequence by the \textit{order} specified during code generation.

\section{Executing WebAssembly in a Database System}
\label{sec:v8}

In the preceding sections, we explained how to compile a QEP and its required libraries to a \Wasm module.  In this
section, we elaborate on how we execute the \Wasm module in an embedded \Wasm engine.  Although this approach works with
any embeddable engine, we describe the process of embedding and executing modules in~V8.

The \Wasm specification requires that each module operates on its personal memory.  This memory is provided by the
engine, here~V8.  To execute a compiled QEP inside the engine, all required data (tables, indexes, etc.) must reside in
the module's memory.  One way to achieve this is by copying all data from the host to the module's memory.  However,
this incurs an unacceptable overhead of copying potentially large amounts of data before executing the QEP.  An
alternative is to use callbacks from the module to the host to transfer single data items on demand.  For such a
purpose, V8 allows for defining functions in the embedder that can be called from the embedded code.  However, such
callbacks also incur a tremendous overhead, because the VM has to convert parameters and the return value from the
representation in embedded code to the representation in the embedder and vice versa.  At the time of writing, V8
provides no method to use pre-allocated host memory as a module's memory.  Therefore, we patch V8 to add a function for
exactly that purpose: \lstinline{SetModuleMemory()} sets the memory of a \Wasm module to a region of the host memory.
While this function enables us to provide a single consecutive memory region from the host to the module, it is not
sufficient to provide multiple tables or indexes (which need not reside in a single consecutive allocation) to a module.
The problem is that \Wasm currently only supports 32\,bit addressing.  Hence, we cannot simply assign the entire host
memory to the module.  Instead, we are limited to 4\,GiB of addressable \emph{linear memory} inside the module.  We work
around this limitation by employing a technique named \emph{rewiring}.

\subsection{Accessing Data by Rewiring}
\label{sec:v8-rewiring}

\begin{figure}
    \centering
    \begin{tikzpicture}[node distance=0pt]
        \tikzstyle{Table}=[block,minimum width=20mm]
        \tikzstyle{Name}=[anchor=south,font=\footnotesize,yshift=-.5ex]
        \tikzstyle{Conn}=[shorten <=6pt,shorten >=6pt,postaction={decorate,decoration={markings,mark=at position .5 with {\arrow{stealth}}}}]
        \tikzstyle{Patt}=[minimum width=20mm-1.2pt,anchor=north,outer sep=0pt]
        \tikzstyle{Addr}=[align=right,font=\ttfamily\tiny,inner sep=0pt]
        \tikzstyle{Memory}=[draw,very thick,rounded corners,inner sep=3pt]
        \tikzstyle{Range}=[|{stealth}-{stealth}|,shorten <=2pt,shorten >=2pt,font=\tiny]

        \node[Table,minimum height=20mm] (vm) {};
        \node[Name,align=center,text width=20mm] (wasm_name) at (vm.north) {Virtual Address Space};
        \node[Patt,below=0mm+1.2pt of vm.north,minimum height=5mm-1.2pt,pattern=north west lines,pattern color=PineGreen!25] {};
        \node[Patt,below=5mm+.6pt of vm.north,minimum height=10mm-.6pt,pattern=north east lines,pattern color=red!25] {};
        \node[Patt,below=15mm+.6pt of vm.north,minimum height=5mm-1.2pt,pattern=crosshatch,pattern color=blue!25] {};
        \node[Addr,left=of vm.north west,anchor=north east] (wasm_start) {0x7ffd\,8000\,0000:};
        \node[Addr,left=of vm.south west,anchor=south east] (wasm_end) {0x7ffe\,7fff\,ffff:};
        \draw[Range,transform canvas={xshift=-4pt}] (wasm_start.south east) -- node[left] {4\,GiB} (wasm_end.north east);

        \node[Table,right=14mm of vm.north east,anchor=north west,yshift=+10mm,minimum height=5mm,pattern=north west lines,pattern color=PineGreen!25] (A) {};
        \node[Name] (name_A) at (A.north) {Table A (1\,GiB)};
        \node[Table,below=5mm of A,minimum height=25mm,pattern=north east lines,pattern color=red!25] (B) {};
        \node[Name] at (B.north) {Table B (5\,GiB)};
        \node[Table,below=5mm of B,minimum height=5mm,pattern=crosshatch,pattern color=blue!25] (rs) {};
        \node[Name] at (rs.north) {Result Set (1\,GiB)};

        \draw[decorate,decoration={brace,mirror,amplitude=4pt,raise=2pt}] (A.north west) -- coordinate (from_A) (A.south west);
        \draw[decorate,decoration={brace,amplitude=4pt,raise=2pt}] (vm.north east) -- coordinate (A_to_wasm) ($(vm.north east)-(0,5mm-1pt)$);
        \draw[Conn] (from_A) -| node[above,font=\tiny] (size_A) {1\,GiB} ($(from_A)!.5!(A_to_wasm)$) |- (A_to_wasm);
        \draw[decorate,decoration={brace,mirror,amplitude=4pt,raise=2pt}] ($(B.north west)-(0,10mm)$) -- coordinate (from_B) ($(B.north west)-(0,20mm)$);
        \draw[decorate,decoration={brace,amplitude=4pt,raise=2pt}] ($(vm.north east)-(0,5mm+1pt)$) -- coordinate (B_to_wasm) ($(vm.north east)-(0,15mm-1pt)$);
        \draw[Conn] (from_B) -| ($(from_B)!.5!(B_to_wasm) + (5pt,0)$) |- coordinate (rewire_B) (B_to_wasm);
        \node[below left=-1pt and -1.5pt of rewire_B,font=\tiny] {2\,GiB};
        \draw[draw=black!60,decorate,decoration={brace,mirror,amplitude=4pt,raise=2pt},dashed,dash pattern=on 1.5pt off 1.5pt] (B.north west) -- coordinate (from_B_old) ($(B.north west)-(0,10mm)$);
        \draw[draw=black!60,shorten <=6pt,dashed,dash pattern=on 1.5pt off 1.5pt] (from_B_old) -| (rewire_B);
        \draw[decorate,decoration={brace,mirror,amplitude=4pt,raise=2pt}] (rs.north west) -- coordinate (from_rs) (rs.south west);
        \draw[decorate,decoration={brace,mirror,amplitude=4pt,raise=2pt}] (vm.south east) -- coordinate (rs_to_wasm) ($(vm.south east)+(0,5mm)$);
        \draw[Conn] (from_rs) -| node[above right,font=\tiny] {1\,GiB} ($(from_rs)!.5!(rs_to_wasm) - (5pt,0)$) |- (rs_to_wasm);

        \coordinate (coord_linmem) at ($(vm.south)-(0,27mm)$);
        \draw[draw=black!15,-{Triangle[width=14mm,length=8mm]},line width=8mm] (vm.south) -- node[midway,rotate=90,font=\tiny,align=center] {\texttt{SetModuleMemory()}\\(our patched V8)} (coord_linmem);

        \node[draw,dashed,minimum width=20mm,minimum height=20mm,anchor=north] (linmem) at (coord_linmem) {};
        \node[font=\tiny,anchor=north] (linmem_name) at (linmem.south) {Module Linear Memory};
        \node[Patt,below=0mm+1.2pt of linmem.north,minimum height=5mm-1.2pt,pattern=north west lines,pattern color=PineGreen!25] (patt_A) {};
        \node[Patt,below=5mm+.6pt of linmem.north,minimum height=10mm-.6pt,pattern=north east lines,pattern color=red!25] (patt_B) {};
        \node[Patt,below=15mm+.6pt of linmem.north,minimum height=5mm-1.2pt,pattern=crosshatch,pattern color=blue!25] (patt_rs) {};
        \node[Addr,left=of linmem.north west,anchor=north east] (linmem_start) {0x0000\,0000:};
        \node[Addr,left=of linmem.south west,anchor=south east] (linmem_end) {0xffff\,ffff:};
        \draw[Range,transform canvas={xshift=-4pt}] (linmem_start.south east) -- node[left] (linmem_size) {4\,GiB} (linmem_end.north east);
        \node[anchor=west,font=\tiny] (linmem_read) at ($(patt_A.east)!.5!(patt_B.east)$) {read};
        \node[right=3pt of patt_rs,font=\tiny] (linmem_write) {write};
        \draw[-,shorten <=-2pt,shorten >=2pt,out=110,in=0] (linmem_read) to (patt_A);
        \draw[-,shorten <=-2pt,shorten >=2pt,out=-110,in=0] (linmem_read) to (patt_B);
        \draw[-,shorten <=-2pt,shorten >=2pt] (linmem_write) -- (patt_rs);

        \begin{pgfonlayer}{bg}

            \node[fit=(linmem)(linmem_name)(linmem_start)(linmem_end)(linmem_size)(linmem_read)(linmem_write),inner sep=0pt] (module_fit) {};
            \node[Memory,fit={([yshift=15pt]module_fit.north west)(module_fit.south east)},inner sep=1pt] (module) {};
            \node[draw,very thick,rounded corners,anchor=north west] at (module.north west) {\textbf{Module}};

            \node[Memory,fit={([yshift=12pt]module.north west)(module.south east)},inner sep=3pt] (v8) {};
            \node[draw,very thick,rounded corners,anchor=north west] at (v8.north west) {\textbf{V8}};

            \node[Memory,fit=(vm)(A)(B)(rs)(name_A)(wasm_start)(wasm_end)(wasm_name)(v8)] (host) {};
            \node[draw,very thick,rounded corners,anchor=north west] at (host.north west) {\textbf{Host Memory}};

            \node[fit=(size_A)(from_A)(A_to_wasm)(from_B)(B_to_wasm)(from_rs)(rs_to_wasm)] (rewiring_fit) {};
            \node[fill=black!15,rounded corners,inner sep=0pt,fit=(rewiring_fit.north)(rewiring_fit.south),minimum width=9.5mm] (rewiring) {};
            \node[font=\tiny,above=-3pt of rewiring] {rewiring};

        \end{pgfonlayer}

        \path (module.north east) -| coordinate (callback_start) (rewiring.south);
        \draw[-{Stealth},very thick,draw=Dandelion!90!black,out=0,in=-90,looseness=1.5] ([yshift=-5mm]callback_start) to node[fill=white,fill opacity=.9,text opacity=1,inner sep=1pt,right=1mm,font=\ttfamily\tiny] {rewire\_next\_chunk(B)} (rewiring.south);

    \end{tikzpicture}
    \vspace*{-8pt}
    \caption{Example of mapping tables and output to a module's memory.  The module can callback to the host to request
    mapping the next 2\,GiB~chunk of table~B.}
    \label{fig:v8-rewiring}
\end{figure}
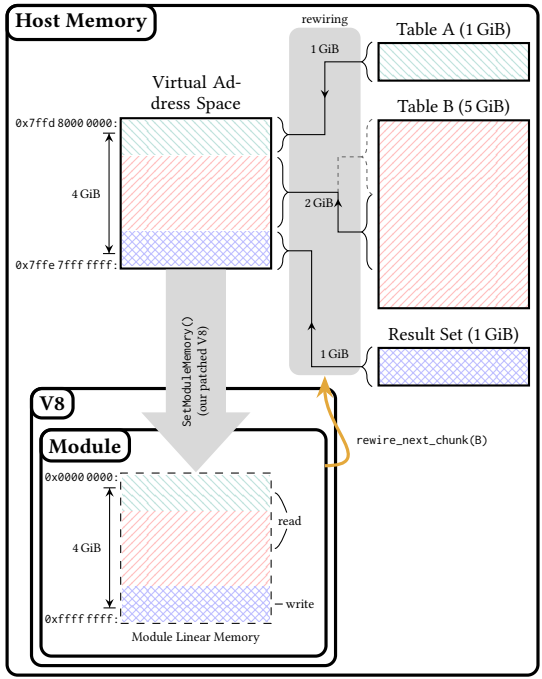

\emph{Rewiring}~\cite{schuhknecht2016ruma} allows for manipulating the mapping of virtual address space to physical
memory from user space.  In particular, it enables us to map the same \emph{physical} memory at two distinct
\emph{virtual} addresses.  We exploit this technique to have data structures residing in distinct allocations appear
consecutively in virtual address space and then use this address range as the module's memory.

We exemplify this technique in \autoref{fig:v8-rewiring}.  Assume a query accessing two tables \texttt{A} and
\texttt{B}.  The tables reside in completely independent memory allocations, hence there is no single 4\,GiB virtual
address range that contains both \texttt{A} and \texttt{B} entirely.  Further, the query computes some results and we
therefore allocate 1\,GiB of memory to store the query's result set.  To give the module access to all required memory,
we first allocate consecutive 4\,GiB in virtual address space.  Then we \emph{rewire} table~\texttt{A}, a portion of
table~\texttt{B}, and the memory for the result set into the freshly allocated virtual memory.  Finally, we call
\lstinline{SetModuleMemory()} with the freshly allocated virtual memory.

The module now has access to both tables and can write its results to the memory allocated for the result set.  Note
that table~\texttt{B} is 5\,GiB and cannot be rewired entirely into the virtual memory for the module.  To give the
module access to the entire table, we install a callback \lstinline{rewire_next_chunk()} that lets the host rewire the
next 2\,GiB chunk of table~\texttt{B} once the module has processed the currently rewired chunk.  This way, the module
can iteratively process entire~\texttt{B} in chunks of~2\,GiB.

\subsection{Result Set Retrieval}

Similarly to how data is made available to the module, we use rewiring to communicate the result set back to the host.
As can be seen in \autoref{fig:v8-rewiring}, the module writes the result set to a rewired allocation of 1\,GiB.  If the
module produces a result set of more than 1\,GiB, it produces the result set in chunks and issues a callback in between
to have the host process the current chunk of results.


\section{Related Work}
\label{sec:related}

We begin our discussion of related work with the vectorized model, then proceed to discuss compilation-based approaches,
before we turn to adaptive query execution.

\textbf{Vectorization.} \ca{graefe94volcano} proposes a unified and extensible interface for the implementation of
relational operators in \noun{Volcano}, named \emph{iterator} interface.  \ca{ailamaki1999time} analyze query execution
on modern CPUs and find that poor data and instruction locality as well as frequent branch misprediction impede the CPU
from processing at peak performance.  \ca{boncz2005monet} identify tuple-at-a-time processing as a limiting factor of
the \noun{Volcano} iterator design, that leads to high interpretation overheads and prohibits data parallel execution.
To overcome these limitations, \citeauthor{boncz2005monet} propose \emph{vectorized} query processing, implemented in
the \noun{X100} query engine within the column-oriented \noun{MonetDB} system.  \ca{menon2017relaxed} build on the
vectorized model and introduce \emph{stages} to dissect pipelines into sequences of operators that can be \emph{fused}.
By fusing operators, \citeauthor{menon2017relaxed} are able to vectorize multiple sequential relational operators.
Their implementation in \noun{Peloton}~\cite{pavlo2017self} shows that operator fusion increases the degree of
inter-tuple parallelism exploited by the CPU.

\textbf{Compilation.}  \ca{rao2006compiled} explore compilation of QEPs to \noun{Java} and having the JVM JIT-compile
and load the generated code.  However, their approach sticks to the \noun{Volcano} iterator model, restricting
compilation from unfolding its full potential.  With \noun{HIQUE}, \ca{krikellas2010generating} propose query
compilation to \Cpp~code by dynamically instantiating operator templates in topological order.  They report query
compilation times in the hundreds of milliseconds.  \ca{neumann2011efficiently} presents compilation of pipelines in the
QEP to tight loops in \noun{LLVM}.  Complex algorithms are implemented in \Cpp and pre-compiled, to be linked with and
used by the compiled query.  With the implementation in \hyper, \citeauthor{neumann2011efficiently} achieves
significantly reduced compilation times in the tens of milliseconds.  \ca{klonatos2014building} address the system
complexity and the associated development effort of compiling query engines in their \noun{LegoBase} system, where
metaprogramming is used to write a query engine in Scala~LMS that when partially evaluated on an input QEP yields
specialized C~code that implements the query.  Despite the clean design, the code generation through partial evaluation
as well as the compilation of the generated code leads to compilation times in the order of seconds.

\textbf{Adaptive.}  Among the most recent advancements in query execution is adaptive execution by
\ca{kohn2018adaptive}, where the QEP is initially executed by interpretation while being compiled to machine code in the
background.  Once compilation completes, the compiled code takes over execution.  Switching execution modes is enabled
by \emph{morsel-wise} query processing~\cite{leis2014morsel}.  This approach, implemented in \noun{HyPer}, shows
promising results as it unites peak performance for long-running queries with low start-up costs for short-lived
queries.  However, this approach comes with the significant drawback of an immense development effort:
\citeauthor{kohn2018adaptive} devise their own LLVM bytecode as well as an interpreter thereof and a compiler to
translate QEPs to said bytecode.  Quite ironically, the authors briefly compare their work to~V8 and consider their work
\say{a database-specific implementation of similar ideas}~\cite{kohn2018adaptive}, yet claim that an automatic solution
like~V8 would fail to achieve competitive performance.  With our work, we hope to convince the reader
otherwise.\done\Todo{Mention Umbra?}

\section{Evaluation}
\label{sec:eval}

We have explained how to compile QEPs to \Wasm and how to perform JIT code generation of library routines.  We motivated
this approach with the ability to specialize the generated code to the actual query to execute.  In this section, we
want to confirm that specialization enables more efficient implementations of QEPs and at the same time code generation
via \Wasm reduces compilation times drastically.  To evaluate the feasibility and profitability of our approach, we
conduct a detailed experimental evaluation.  We begin by evaluating the performance of QEP building blocks, then we look
at TPC-H queries, before looking closer at compilation times.

\subsection{Experimental Setup}
\label{sec:eval-setup}

We implement our approach in \mutable~\cite{mutable}, a main-memory database system currently developed at our group.
Incoming SQL queries are translated to a graph representation, where unnesting and decorrelation is performed as far as
possible, similar to the approach described by \ca{neumann2015unnesting}.  The optimizer of \mutable computes an optimal
join order for the query and constructs the QEP, that is then passed to the \Wasm backend of \mutable, where it is
translated to a \Wasm module and dispatched to the \Wasm platform for execution.  Although \mutable supports arbitrary
data layouts, we conduct all experiments using a columnar layout.  Since \mutable does not yet support multi-threading,
all queries run on a single core.

We compare to three systems: (1)~\postgres~13.1 as representative for \noun{Volcano}-style tuple-at-a-time processing,
(2)~\duckdb~v0.2.3, implementing the vectorized model as in~MonetDB/X100, and (3)~\hyper, an adaptive system performing
interpretation and compilation of LLVM bytecode, as provided by the \texttt{tableauhyperapi} \textsc{Python} package in
version~0.0.11952.  For \postgres, we disable JIT compilation as it does not improve execution time in any of our
experiments.\done\todo{Elaborate.}  Because our version of \hyper uses the adaptive approach from \ca{kohn2018adaptive},
which cannot be disabled, we cannot distinguish between compilation and execution times.  We run all our experiments on
a machine with an AMD Ryzen Threadripper 1900X CPU with 8~physical cores at 3.60~GHz and 32~GiB main memory.  All data
accessed in the experiments is memory resident.  We repeat each experiment five times and report the median.

\begin{figure*}
    \centering
    \vspace*{-11pt}
    \includegraphics[height=8pt]{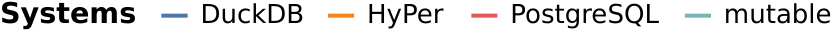}
    \vspace*{-1em}
\end{figure*}

\subsection{Performance of Query Building Blocks}
\label{sec:eval-operators}

With our first set of experiments, we evaluate the performance of individual query building blocks across different
systems.  We use a generated data set with multiple tables and 10~million rows per table.  Tables contain only integer
and floating-point columns, where integer values are chosen uniformly at random from the entire integer domain and
floating-point values are chosen uniformly at random from the range $[0; 1]$.  All data is shuffled and all columns are
pairwise independent.  For \mutable, we report execution time as not including compilation time.  Further, if not
mentioned otherwise, we enforce compilation with the optimizing \noun{TurboFan} compiler.  For \hyper, we report the
end-to-end execution time, which may include time spent on compilation.

\begin{figure}
    \centering
    \includegraphics[scale=.65]{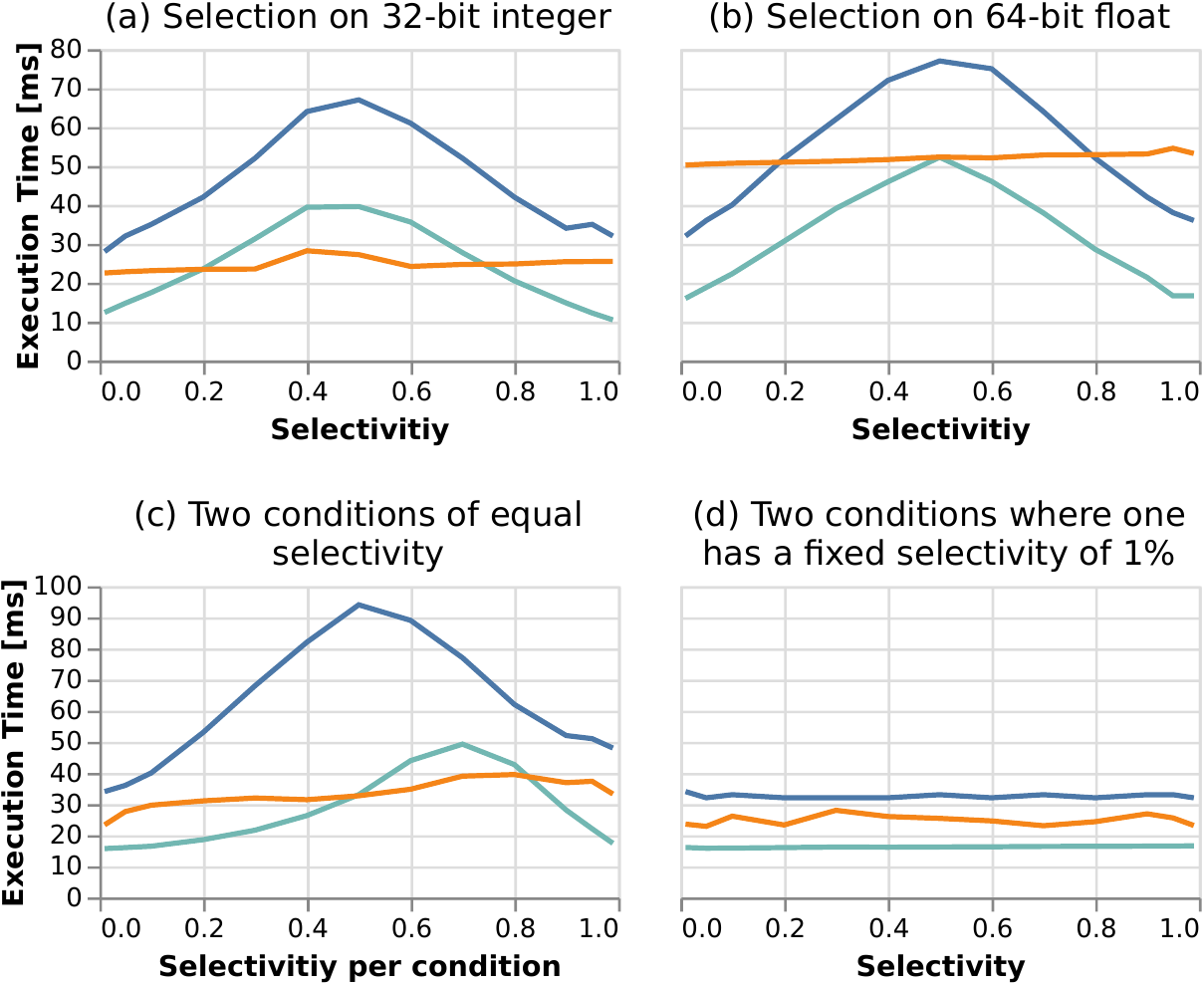}
    \vspace*{-10pt}
    \caption{Evaluation of selection with one and two one-sided range predicates.}
    \label{fig:eval-sel}
\end{figure}

\textbf{Selection.}  In our first experiment, we evaluate the performance of selection with the query
\lstinline[language=sql]{SELECT COUNT(*) FROM T WHERE T.x < X;} and vary \lstinline{X} to achieve different
selectivities.  \autoref{fig:eval-sel}~(a) and~(b) show our measurements for selection on a $32$-bit integer and a
$64$-bit floating-point column, respectively.  We omit our findings for \postgres, as the times are over $200$\,ms.  The
charts show that the execution times of both \mutable and \duckdb depend on the selectivity of the selection.  At
selectivities around~$50\%$, frequent branch misprediction causes performance to
deteriorate~\cite{sompolski2011vectorization, ross2002conjunctive}.  With selectivities closer to $0\%$ or $100\%$, the
frequency of branch misprediction declines and performance improves.  The execution time of \hyper remains unaffected by
varying selectivity; we assume that \hyper compiles branch-free code.  We can see that \mutable outperforms \duckdb on
all selectivities and for both integer and floating-point columns.  This is likely the case because \duckdb, which
implements the vectorized execution model, has the overhead of maintaining a selection
vector~\cite{sompolski2011vectorization, pirk2016voodoo}.  For the integer column, \hyper outperforms \mutable at
selectivities from $20$\% to $75$\%, outside this range \mutable is up to $2x$~faster than \hyper.  For the
floating-point column, \mutable outperforms \hyper on all selectivities, with a speedup of up to~$2.5x$.

We conduct two additional experiments, where we perform a selection on two independent integer columns with the query
\lstinline[language=sql]{SELECT COUNT(*) FROM T WHERE T.x < X AND T.y < Y}.  In the first experiment, shown in
\autoref{fig:eval-sel}~(c), \lstinline{X} and \lstinline{Y} are both varied with equal selectivity.  This means, the
overall selectivity of the selection is the squared selectivity of either condition.  Since \mutable does not implement
short-circuit evaluation and instead evaluates the selection as a whole, a selectivity of $\sqrt{50\%} \approx 71\%$ per
condition presents the worst-case for branch prediction with a time of $50$\,ms.  \duckdb, which implements the
vectorized model, must first evaluate one condition to a selection vector before evaluating the second condition on the
selected rows.  Because the conditions are evaluated individually, branch misprediction occurs up to twice as often and
branch prediction is worst at a selectivity of $50\%$ with an execution time around $90$\,ms.  As the selectivity grows,
the second condition must be evaluated more often.  This can be seen in the slight asymmetry in execution times, where a
selectivity close to $100\%$ takes around $50$\,ms and a selectivity close to $0\%$ takes less than $40$\,ms.  \hyper's
execution time slightly grows with the selectivity from around $30$\,ms at $0\%$ to $40$\,ms at $100\%$.  We assume that
\hyper again produces branch-free code.  However, the value required in the second condition might only be loaded if the
first condition is satisfied, explaining the slight growth in time.

In the second experiment, shown in \autoref{fig:eval-sel}~(d), \lstinline{X}~is varied while~\lstinline{Y} is fixed such
that the right condition has a selectivity of~$1$\%.  The overall selectivity of the selection is hence in the range
from~$0\%$ to~$1\%$.  Since \mutable evaluates the entire selection as a whole, branch prediction works reliably well
and we observe a constant execution time of around $15$\,ms.  \duckdb likely evaluates the more selective condition
first, resulting in a constant execution time around~31\,ms.  As \hyper's execution time is independent of the
selectivity, the execution time remains stable around $25$\,ms, similar to~(c).

\begin{figure}
    \centering
    \includegraphics[scale=.65]{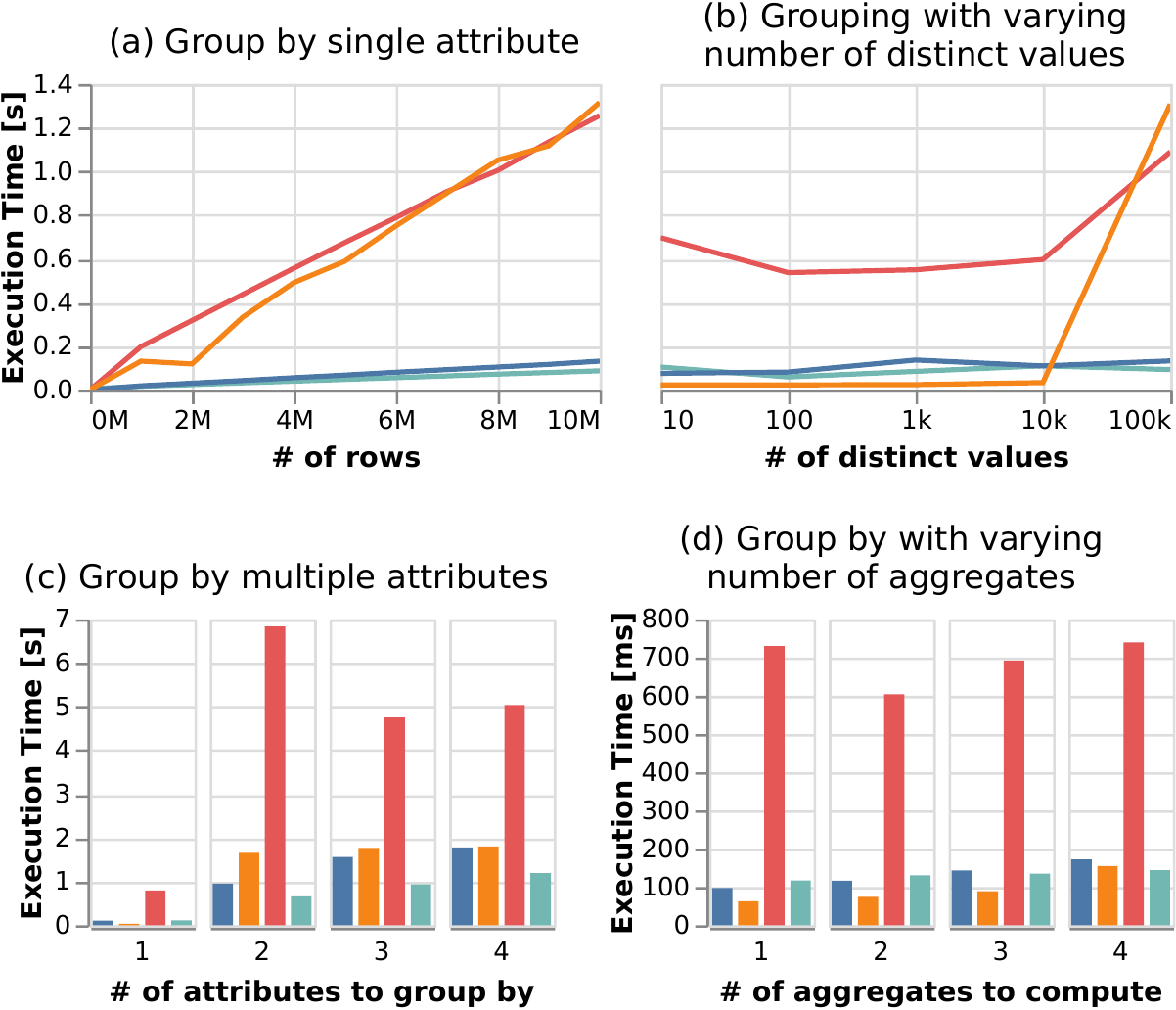}
    \vspace*{-10pt}
    \caption{Evaluation of grouping~\&~aggregation.}
    \label{fig:eval-group_by}
\end{figure}

\textbf{Grouping~\&~Aggregation.}  Our next experiment evaluates the performance of grouping and aggregation.  We
vary the experiment in several dimensions: the number of rows in the table, the number of distinct values in the column
being grouped by, and the number of attributes to group by.  Our first query is
\lstinline[language=sql]{SELECT COUNT(*) FROM (SELECT 1 FROM T GROUP BY T.x) AS U}
which computes the number of distinct values in column~\lstinline{T.x}, which is $100$k here.  We show our findings in
\autoref{fig:eval-group_by}~(a).  We can see that \postgres and \hyper take relatively long to evaluate the query in
comparison to \duckdb and \mutable.  For \hyper, we assume that the compiled QEP is linked with a pre-compiled hash
table implementation, leading to significant function call overheads for table lookups.  For \postgres we can only
assume that the implementation of hash-based grouping is relatively slow.  \duckdb can rely on an efficient
implementation of hash-based grouping by a single column.  \mutable generates a hash table implementation specialized
for the column being grouped by.  \mutable slightly outperforms \duckdb with $110$\,ms versus $131$\,ms at $10$M rows.

We reuse the above query, yet this time we leave the number of rows fixed at 10M and vary the number of distinct values
in the column being grouped by.  The number of distinct values directly corresponds to the number of entries in the hash
table used for grouping.  Our results are shown in \autoref{fig:eval-group_by}~(b).  \postgres still performs relatively
poor in comparison to the other systems, while \duckdb and \mutable perform consistently well.  Very interesting are the
observations for \hyper: up until 10k distinct values, \hyper's execution time is close to zero.  We assume that the
system can answer the query from internal statistics rather than actually executing the query.  Once the number of
distinct values grows too large, it appears that \hyper cannot rely on statistics anymore and the query must be
executed.  In the case of 100k distinct values, \hyper's execution time is actually the highest of all systems.  Similar
to \autoref{fig:eval-group_by}~(a), \mutable and \duckdb lie very close to each other with \mutable outperforming
\duckdb in four out of five cases.

In our next variation of the experiment, we vary the number of attributes to group by from one to four.  The attributes
are chosen such that at least $10$k groups are formed.  We present our findings in \autoref{fig:eval-group_by}~(c).  We
can see that all systems' execution times spike when increasing the number of attributes to group by from one to two.
This behaviour can be explained as follows: when grouping by a single attribute, a hash can be computed directly from
the attribute's value but when grouping by multiple attributes, a hash for the combined attribute values must be
computed, which can become significantly more complex.  When increasing the number of attributes to group by further,
the execution times still increase yet at a smaller rate.  The only system that does not fit into this scheme is
\postgres, with its highest execution time at two attributes.  With two or more attributes to group by, \mutable is the
fastest of the systems, outperforming the others by at least $1.5x$.  We credit this behaviour to \mutable's generation
of specialized code to compute a hash as well as a specialized hash table, minimizing any overhead for hash computation
or table lookup.

Lastly, we evaluate the performance of aggregation with the query
\lstinline[language=sql]|SELECT MIN(T.y1), ..., MIN(T.yn) FROM T GROUP BY T.x| and vary the number of aggregates to
compute.  Our findings are shown in \autoref{fig:eval-group_by}~(d).  Surprisingly, \postgres's execution times above
$600$\,ms are significantly larger than those of the other systems.  \hyper outperforms the other systems on~1 to~3
aggregates while for 4~aggregates, \mutable performs best.  It is important to see that the slope of the execution time
over the number of aggregates to compute grows least for \mutable.  Hence, \mutable eventually takes the lead at
4~aggregates.

\begin{figure*}[t]
    \centering
    \vspace*{-11pt}
    \includegraphics[height=8pt]{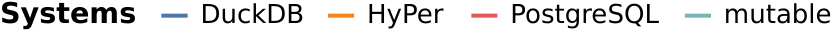}
    \vspace*{-1em}
\end{figure*}

\begin{figure}
    \centering
    \includegraphics[scale=.65]{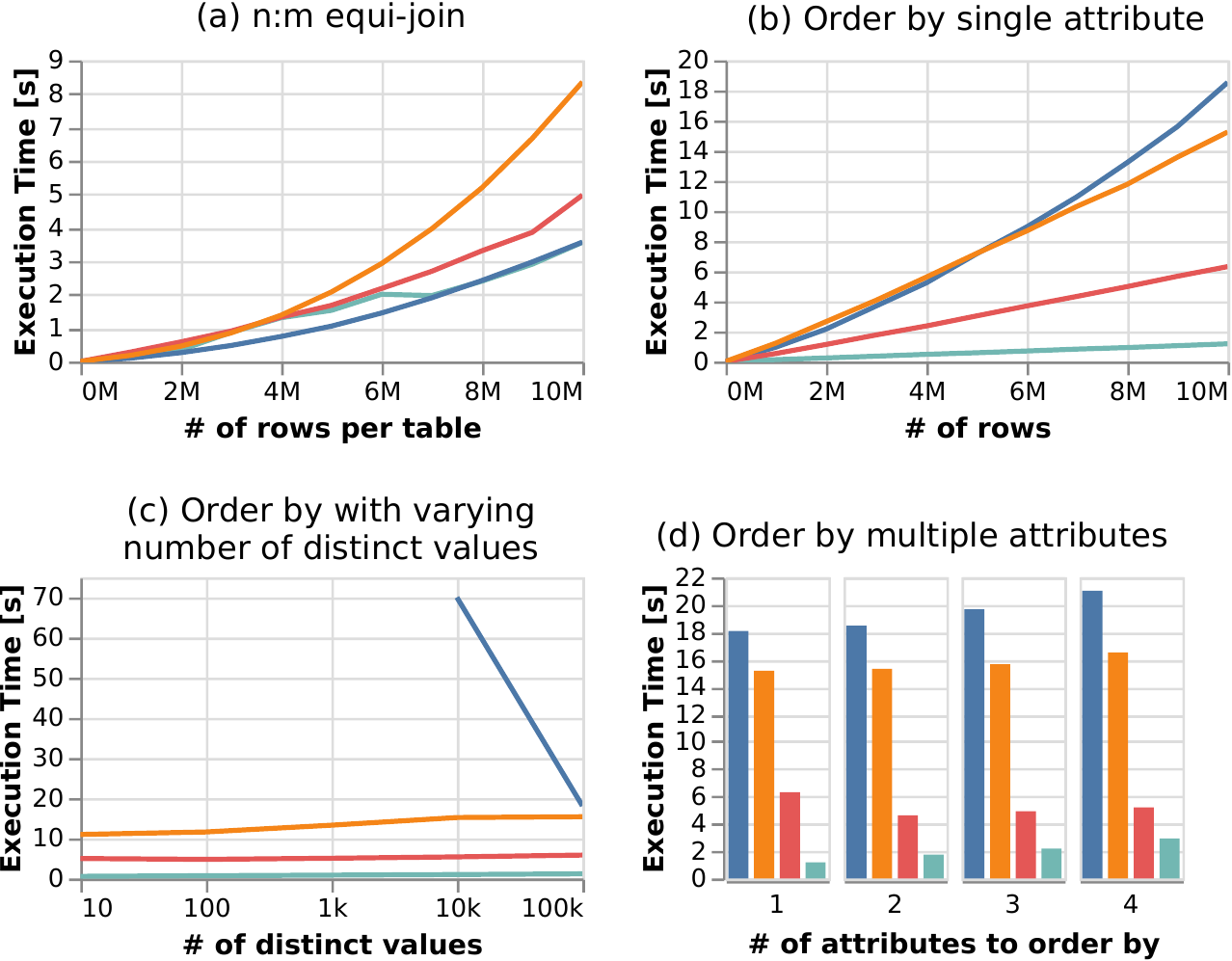}
    \vspace*{-18pt}
    \caption{Evaluation of equi-join and ordering.}
    \label{fig:eval-join_and_order_by}
\end{figure}

\textbf{Equi-Join.}  In this experiment, we evaluate the performance of an $n$:$m$~equi-join.  We perform the join on
non-key columns to avoid the systems using a pre-built index, since \mutable does not yet support indices\footnote{In
particular, \mutable cannot map non-consecutive data structures like indices from process memory into the \Wasm~VM.
This is future work.}.  We present our findings in \autoref{fig:eval-join_and_order_by}~(a).  In the experiment, we
leave the selectivity of the join fixed at $10^{-6}$ and vary the size of the input relations.  All systems show the
expected quadratic curve.  For up to $3$M rows, \postgres is the slowest system.  For more than $3$M rows, \hyper
becomes the slowest of all systems because of its strong quadratic curvature.  \postgres, \duckdb, and \mutable show
very similar performance, with \mutable being slightly slower than \duckdb for less than 7M rows.

\textbf{Sorting.}  Our last experiment evaluates the performance of sorting, as needed in
\lstinline[language=sql]|ORDER BY|-clauses or for merge-join.  Similar to the experiment on grouping, we vary the
experiment in several dimensions: the number of rows in the table, the number of attributes to order by, and the number
of distinct values in the column to order by.  \autoref{fig:eval-join_and_order_by}~(b) to~(d) present our findings.  In
\autoref{fig:eval-join_and_order_by}~(c), we restrict \duckdb to $\ge$10k distinct values as \duckdb's implementation
of \noun{QuickSort} exhibits quadratic running time for almost sorted data.\done\todo{Elaborate \duckdb outside range.}

Throughout all experiments, \mutable significantly outperforms the other systems, with factors up to $4x$.  We credit
this immense performance improvement to our ad-hoc code generation and consequent holistic optimization of the sorting
operation, as described in detail in \autoref{sec:codegen-example}.  \done\todo{Write down observations of evaluating
sorting.  Potentially summarize all three experiments in one paragraph.}

\textbf{Summary.}  With our set of experiments we are able to show that our approach of compiling QEPs to \Wasm not
only provides competitive performance but in many cases improves performance significantly.  We credit these performance
improvements to the ad-hoc code generation of specialized library code and the potential for holistic optimization
by~V8.

\subsection{TPC-H}
\label{sec:eval-tpc_h}

\begin{figure}
    \centering
    \begin{subfigure}[b]{\linewidth}
        \centering
        \includegraphics[scale=.65]{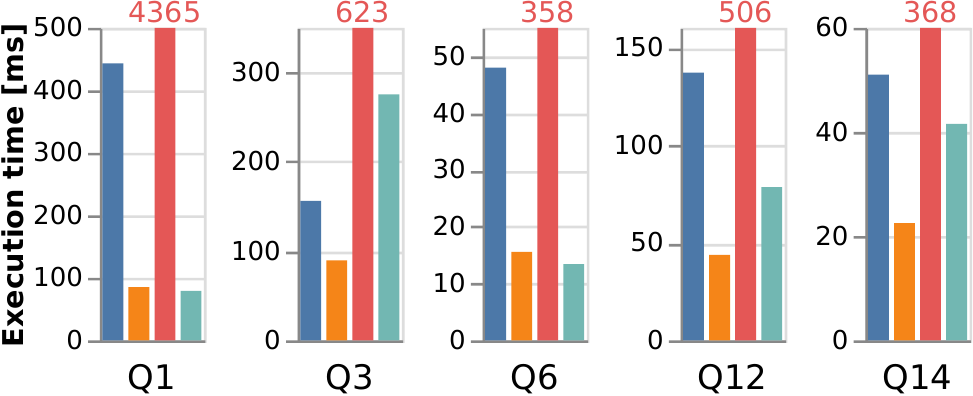}
        \vspace*{-5pt}
        \caption{Execution times.}
    \end{subfigure}

    \begin{subfigure}[b]{.45\linewidth}
        \centering
        \includegraphics[scale=.65]{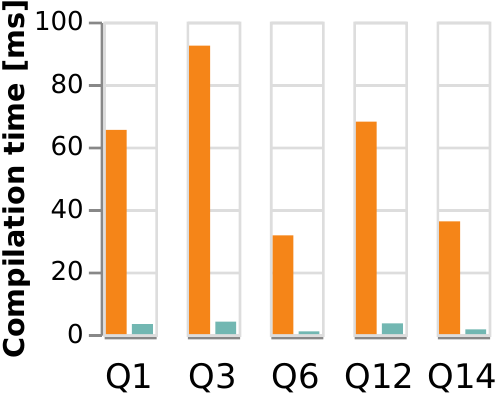}\vspace*{4.2mm}
        \vspace*{-5pt}
        \caption{Compilation times.}
    \end{subfigure}%
    \begin{subfigure}[b]{.55\linewidth}
        \centering
        \includegraphics[scale=.65]{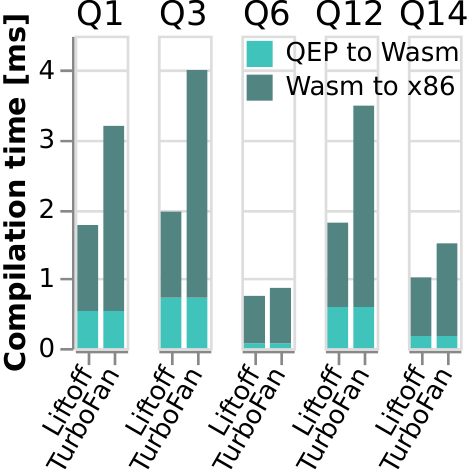}
        \vspace*{-5pt}
        \caption{\mutable's compilation times.}
    \end{subfigure}
    \vspace*{-20pt}
    \caption{Evaluation of TPC-H queries on a SF\,1 database.}
    \label{fig:eval-tpc_h}
\end{figure}

So far, our experiments only focus on individual query building blocks.  Next, we conduct an experimental evaluation of
TPC-H queries.  By the time of writing, \mutable\ -- and in particular our \Wasm backend -- only supports a subset of
SQL and hence we are only able to evaluate the TPC-H queries Q1, Q3, Q6, Q12, and Q14.  For \hyper, we report the
end-to-end time measured as explained in \autoref{sec:eval-setup} as well as the compilation time as reported by the
\hyper WebInterface~\cite{hyper-web}.  For \mutable, we provide detailed timings for the translation of the QEP to \Wasm
and the compilation and execution of \Wasm with \noun{Liftoff} and \noun{TurboFan}.
We present our findings in \autoref{fig:eval-tpc_h}.

For queries Q1 and Q6, \mutable outperforms the other systems while for Q3, Q12, and Q14 \hyper is almost $2x$ faster
than \mutable.  We credit this performance gap to \hyper's superior performance on foreign-key joins.  With regard to
compilation times, \mutable's pipeline with the optimizing \noun{TurboFan} compiler is up $36x$ faster than \hyper's
LLVM-based compilation pipeline.  At the same time, for Q1 and Q6, \mutable is able to outperform \hyper's adaptive
execution.  With the \noun{Liftoff} compilation pipeline, we are able to push down compilation times further at the
sacrifice of execution speed.  Our results confirm that the compilation of QEPs to \Wasm is indeed competitive in terms
of execution speed to a fully compiling and optimizing pipeline, like \noun{LLVM} in \hyper.

\section{Conclusion}
\label{sec:conclusion}

In this work, we explored execution of QEPs in a database system by compilation to \Wasm.  With our approach, we are
able to achieve compilation times under one millisecond even for complex queries and by that we were able to remedy the
recurring point of criticism that compilation-based approaches impose a high latency to query execution.  Further, we
have shown that our approach yields highly efficient machine code without the side effect of high compilation times.  By
relying on a system particularly designed for JIT compilation like V8 and a portable, low-level language like \Wasm, we
are able to lift a burden from the shoulders of database engineers.\done\todo{This is again very similar to LLVM!}

We are convinced that our approach is considerably simpler to understand and implement than current state-of-the-art
solutions.  By relying on successful, battle-tested infrastructure for JIT compilation and execution, we significantly
reduce the required development effort to build an adaptive yet highly efficient query execution engine.  With the
ongoing standardization of \Wasm~\cite{WasmMDN,Wasm} and the immense interest and amount of ongoing work in engines
supporting this language~\cite{V82008,SpiderMonkey,Wasmtime,Wasmer,WebKit}, our approach provides a reliable and
future-proof solution to adaptive query execution.

\bibliographystyle{ACM-Reference-Format}
\bibliography{main}

\clearpage
\appendix
\appendixpage

\section{\Wasm Complete Example}
\label{apx:wasm-example}

The short example in \autoref{sec:wasm-example} focuses on the selection $\sigma_\text{R.val < 3.14}$.  Below, we show a
full example including a table scan and projection, that is constructed by compiling the query
\lstinline[language=sql]{SELECT 1 FROM R WHERE R.val < 3.14} to \Wasm.

\begin{figure}[H]
\begin{lstlisting}[
        language=WebAssembly,
        mathescape=false,
        basicstyle=\ttfamily\fontsize{6}{6}\selectfont
    ]
(module
 (type $i32_=>_i32 (func (param i32) (result i32)))
 (import "env" "head_of_heap" (global $head_of_heap i32))
 (import "env" "R_num_rows" (global $R_num_rows i32))
 (import "env" "R.val" (global $R.val i32))
 (memory $0 1 65535)
 (export "memory" (memory $0))
 (export "run" (func $run))
 (func $run (; 1 ;) (param $0 i32) (result i32)
  (local $1 i32)
  (local $2 i32)
  (local $3 i32)
  (local $4 i32)
  (block $run.body_0 (result i32)
   (local.set $1 (global.get $head_of_heap))
   (local.set $4 (global.get $R.val))
   (block $pipeline.scan_R_1
    (if
     (i32.lt_u
      (local.get $3)
      (global.get $R_num_rows)
     )
     (loop $scan_R_2
      (block $scan_R_2.body_3
       (if
        (f64.lt
         (f64.promote_f32
          (f32.load
           (local.get $4)
          )
         )
         (f64.const 3.14)
        )
        (block $filter.accept_4
         (i32.store
          (local.get $1)
          (i32.const 1)
         )
         (local.set $1
          (i32.add
           (local.get $1)
           (i32.const 8)
          )
         )
         (local.set $2
          (i32.add
           (local.get $2)
           (i32.const 1)
          )
         )
        )
       )
       (local.set $4
        (i32.add
         (local.get $4)
         (i32.const 4)
        )
       )
       (local.set $3
        (i32.add
         (local.get $3)
         (i32.const 1)
        )
       )
       (br_if $scan_R_2
        (i32.lt_u
         (local.get $3)
         (global.get $R_num_rows)
        )
       )
      )
     )
    )
   )
   (i32.store
    (local.get $1)
    (local.get $2)
   )
   (local.get $1)
  )
 )
)
\end{lstlisting}
    \caption*{Entire \Wasm code for the example in \autoref{sec:wasm-example}.  The bytecode is only $219$~bytes.  The
    variable \texttt{\$1} from \autoref{sec:wasm-example} is here named \texttt{\$4}.}
\end{figure}

\todos

\end{document}